\documentclass[lettersize,journal]{IEEEtran}

\usepackage[dvipsnames]{xcolor}

\usepackage{amsmath,amsfonts}
\usepackage{algorithmic}
\usepackage{algorithm}
\usepackage{array}
\usepackage[caption=false,font=normalsize,labelfont=sf,textfont=sf]{subfig}
\usepackage{textcomp}
\usepackage{stfloats}
\usepackage{url}
\usepackage{verbatim}
\usepackage{graphicx}
\usepackage{cite}
\usepackage{soul}
\usepackage{xspace}
\usepackage{comment}
\usepackage{tabularx}
\usepackage{multirow}
\usepackage{listings}
\usepackage{hyperref}
\usepackage{amssymb}
\usepackage{adjustbox}
\usepackage{enumitem}
\usepackage[hang,flushmargin]{footmisc}
\usepackage[normalem]{ulem}
\usepackage{cleveref}
\usepackage{tabu}
\usepackage{booktabs}
\usepackage{mathtools}
\usepackage{lipsum}
\usepackage{placeins}
\usepackage{mathtools} \usepackage{xspace} 
\newcommand{\FLIP}{\protect\reflectbox{F}LIP\xspace}

\definecolor{codeGreen}{rgb}{0,0.6,0}
\definecolor{codeBlue}{rgb}{0,0,1}
\definecolor{codeRed}{rgb}{0.65,0.11,0.36}
\definecolor{codeGray}{rgb}{0.5,0.5,0.5}
\definecolor{codeMauve}{rgb}{0.58,0,0.82}
\definecolor{codeCyan}{rgb}{0,0.52,0.70}

\def\lstfontsize{\fontsize{7}{7}\selectfont}
\newcolumntype{M}[1]{>{\centering\arraybackslash}m{#1}}

\newcommand{\posthoc}{\textit{post hoc}\xspace}
\newcommand{\Posthoc}{\textit{Post hoc}\xspace}

\newcommand{\insitu}{\textit{in situ}\xspace}
\newcommand{\Insitu}{\textit{In situ}\xspace}
\newcommand{\InSitu}{\textit{In Situ}\xspace}

\newcommand{\tcnn}{Tiny-CUDA-NN\xspace}
\newcommand{\tthresh}{{TTHRESH}\xspace}
\newcommand{\szfamily}{{SZ}\xspace}
\newcommand{\szthree}{{SZ3}\xspace}
\newcommand{\zfp}{{ZFP}\xspace}
\newcommand{\zstd}{{ZSTD}\xspace}
\newcommand{\sperr}{{SPERR}\xspace}
\newcommand{\diva}{DIVA\xspace}

\newcommand{\eg}{e.g.,\xspace}
\newcommand{\ie}{i.e.,\xspace}

\newcommand{\etal}{et al.\xspace}

\newcommand{\tvcg}[1]{{#1}}

\newcommand{\tvcgm}[1]{{#1}}

\usepackage{etoolbox}
\makeatletter
\patchcmd{\@makecaption}
  {\scshape}
  {}
  {}
  {}
\makeatother

\begin{document}

\title{Distributed Neural Representation for\\Reactive \InSitu Visualization}

\author{Qi Wu, Joseph A. Insley, Victor A. Mateevitsi, Silvio Rizzi, Michael E. Papka, and Kwan-Liu Ma
\thanks{Qi Wu and Kwan-Liu Ma are affiliated with the University of California, Davis, USA. Email: \{qadwu\,$|$\,klma\}@ucdavis.edu}
\thanks{Victor A. Mateevitsi and Silvio Rizzi are with Argonne National Laboratory, USA. Email: \{vmateevitsi\,$|$\,srizzi\}@anl.gov}
\thanks{Michael E. Papka is affiliated with Argonne National Laboratory, USA, and the University of Illinois at Chicago, USA. Email: papka@anl.gov}
\thanks{Joseph A. Insley is affiliated with Argonne National Laboratory, USA, and Northern Illinois University, USA. Email: insley@anl.gov}
}

\markboth{Journal of \LaTeX\ Class Files,~Vol.~14, No.~8, August~2021}%
{Qi Wu \MakeLowercase{\textit{et al.}}: Distributed Neural Representation for Reactive \InSitu Visualization}

\maketitle

\begin{abstract}
Implicit neural representations (INRs) have emerged as a powerful tool for compressing large-scale volume data. This opens up new possibilities for \insitu visualization. However, the efficient application of INRs to distributed data remains an underexplored area. In this work, we develop a distributed volumetric neural representation and optimize it for \insitu visualization. Our technique eliminates  data exchanges between processes, achieving state-of-the-art compression speed, quality and ratios. Our technique also enables the implementation of an efficient  strategy for caching large-scale simulation data in high temporal frequencies, further facilitating the use of reactive \insitu visualization in a wider range of scientific problems. We integrate this system with the Ascent infrastructure and evaluate its performance and usability using real-world simulations.
\end{abstract}


\section{Introduction}
\IEEEPARstart{R}{ecent} advances in volume compression have highlighted the effective use of neural networks to implicitly represent volume data~\cite{lu2021compressive}. Such a neural representation offers several advantages. Firstly, it allows for significant reductions in data size by several orders of magnitude while preserving high-frequency details of the data. Secondly, it permits retrieving data values without the need for decompression. Thirdly, it enables access to spatial locations at arbitrary resolutions. The latest developments in this field have also enabled lightning-fast training and high-fidelity interactive volume rendering of neural representations~\cite{wu2022instant}. These advantages make implicit neural representation (INR) a promising technique for handling large-scale volume data.

In practice, a substantial amount of large-scale volume data originates from state-of-the-art scientific simulations running on high-performance computing (HPC) platforms. These simulations demand the efficient distribution of data across numerous processing units to maximize computational power and memory space utilization. This optimization is crucial for enhancing scalability and efficiency, particularly in conducting calculations and communication operations integral to the simulation. However, such prioritization makes the application of INR difficult using popular machine learning pipelines like PyTorch. 
Having a highly efficient INR training system, specifically tailored for data-distributed HPC simulations, is crucial but has yet to be fully explored in the literature. This work addresses this gap by introducing a distributed volumetric neural representation (DVNR) via model parallelism and an efficient machine learning pipeline fully compatible with the computing resource constraints imposed by the simulation.

In this work, we extend a preliminary design of DVNR~\cite{9966405}, aiming to significantly improve its scalability, compression ratio, and quality. This enhancement is achieved by providing a highly optimized CUDA backend and by enabling adaptive parameter tuning, boundary loss, and model compression. 

Our extended design is also motivated by the need for adaptive \insitu visualization. \Insitu visualization performs rendering and data reduction concurrently with the simulation, thereby circumventing the need for costly data I/O. To improve efficiency, modern \insitu visualization infrastructures often allow tasks to be scheduled around data-driven triggers~\cite{InSituTrigger} and programmed using reactive languages like \diva~\cite{wu2020diva}. In studies like causality analysis, critical insights are often embedded in the data preceding a detectable event. Thus, it is necessary to have simulation data temporarily cached  in memory for subsequent visualization and analysis.  However, implementing such a caching mechanism can be challenging due to the sheer size of the simulation data. To tackle this challenge, we further optimize our training pipeline through weight caching, and then use DVNR to create an efficient \insitu temporal caching support for the \diva reactive programming system.

Finally, we integrate DVNR and \diva with the Ascent \insitu visualization and analysis infrastructure~\cite{larsen2022ascent}, featuring seamless interoperations between DVNR and traditional visualization algorithms through reactive programming. Our integration enables the use of DVNR as well as reactive programming in a wider spectrum of real-world, production-ready \insitu scientific simulations.

\tvcgm{We release the source code of our implementation at \url{https://github.com/VIDILabs/distributedvnr}.}
The key contributions of this work can be summarized as follows:
\begin{itemize}
  \item The introduction of a distributed volumetric neural representation (DVNR) with an efficient training system that eliminates the need for interprocess communication.
  \item The optimization of DVNR for HPC environments by simultaneously employing adaptive parameter tuning, boundary loss, model compression, and weight caching.
  \item The implementation of an efficient DVNR-based temporal data caching strategy for reactive \insitu visualization.
  \item The integration of DVNR with production visualization infrastructures for supporting real-world simulations.
\end{itemize}

\section{Related Work}

In this section, we explore the research related to our work. We start with an overview of scientific data compression, followed by brief reviews of model compression and meta-learning for neural networks. Lastly, we summarize recent progress in \insitu visualization.

\subsection{Scientific Data Compression}

Achieving high-quality volume visualization, particularly with large, high-resolution datasets, presents a significant challenge. Lossless algorithms, like Zlib and Zstandard~\cite{Zstandard}, maintain complete data integrity, but typically fall short in achieving high compression ratios beause they rely on discovering repeating byte patterns whereas scientific data is often composed of diverse floating-point numbers. In contrast, lossy algorithms can yield substantially higher compression ratios, accepting some level of information loss as a trade-off. Popular lossy algorithms for scientific data generally fit into three categories: transformation-based, tensor decomposition-based approaches, and prediction-based.

Transformation-based methods use linear transforms to enhance data sparsity, encoding the resultant coefficients to compress efficiently. A notable example is \zfp~\cite{lindstrom2014fixed}, which emphasizes fast random access and minimal errors. \tvcg{\sperr is a more recent example that uses wavelet transforms, {SPECK} coding~\cite{pearlman2004efficient, tang2006three}, and a custom outlier coding algorithm to achieve high-performance compression for structured scientific data~\cite{li2023lossy}.}
Conversely, tensor decomposition methods leverage adaptive bases for transformation. Such strategies include {TAMRESH}~\cite{suter2013tamresh} and \tthresh~\cite{ballester2016lossy, ballester2019tthresh}, which focus on increasing transform-domain sparsity at the expense of having to storing its learned bases.
Prediction-based approaches use predictors
to estimate unknown data points from known values at cube vertices. 
The \szfamily~\cite{tao2017significantly, liang2018error, zhao2021optimizing} is a family of prediction-based compressors capable of adaptively selecting the most suitable prediction method for each dataset. 

Deep learning has also been explored to advance volume data compression. Jain~\etal~\cite{jain2017compressed} introduced an encoder-decoder network designed for time-variable multivariate volumes. Lu~\etal~\cite{lu2021compressive} combined SIREN-based INRs~\cite{sitzmann2020implicit} and ResNet~\cite{he2016deep} for compressing scientific volume data. Subsequent enhancements by fV-SRN~\cite{weiss2021fast} and Wu~\etal~\cite{wu2022instant} utilized grid-based encoding to significantly speed up both training and inference, with Wu~\etal's technique also facilitating the first instance of real-time INR visualization via wavefront rendering algorithms. Their work lays the foundation for this project. Wurster~\etal's APMGSRN~\cite{wurster2023adaptively} further enhanced grid-based encoding by incorporating adaptively positioned multi-grids. For time-varying volumes, Han~\etal~\cite{han2022coordnet} proposed a novel ResNet-based INR framework. For handling high-resolution sparse volumes, Doyub~\etal~\cite{kim2022neuralvdb} combined INRs with the OpenVDB framework.
Recently, we introduced a methodology that enables \insitu generation of INRs for distributed volume data using model parallelism, initially presented in a poster~\cite{9966405}. This paper expands on our preliminary work, delving into performance optimizations and adaptations tailored for reactive \insitu visualization.

\tvcgm{The concept of representing the volumetric domain with independent INRs was introduced by DeRF~\cite{rebain2021derf} and KiloNeRF~\cite{reiser2021kilonerf}, adopted by APMGSRN~\cite{wurster2023adaptively}, and further refined by MINER~\cite{saragadam2022miner} and ECNR~\cite{tang2024ecnr} using a hierarchical partitioning strategy. However, these works assume the entire volume is available at once and use model parallelism solely for acceleration. In contrast, our work focuses on distributed data that is already partitioned by the simulation and computed \insitu. We employ model parallelism to minimize data communication costs, whereas performing a hierarchical re-partitioning of the data, as done by MINER and ECNR, would be too costly. Compared to APMGSRN, our method prioritizes overall computation speeds, utilizing a more efficient INR~\cite{muller2022instant} along with compatible training~\cite{tcnn} and visualization systems~\cite{wu2022instant}.}

\subsection{Model Compression}

While INRs already provide compact representations of large volume data, the underlying neural networks can be further compressed to enhance the compression ratio. Popular methods for reducing model size include pruning, which increases data sparsity~\cite{han2015deep, han2015learning, frankle2018lottery, gale2019state}, and quantization, which dynamically minimizes the bit requirement for storing weights~\cite{han2015deep, lin2017towards, tung2018clip}. For SIREN-based INRs, Lu~\etal~\cite{lu2021compressive} adopts a model compression method that clusters the weights of hidden layers using K-means before quantization. However, their method can only compress MLP layers. For grid encoding-based INRs, Girish~\etal~\cite{girish2023shacira} concurrently introduced a feature grid compression method using reparameterized latents with entropy regularization, although at the cost of longer model training times. Model compression can also be achieved through knowledge distillation, however, this is beyond the scope of this paper. A more comprehensive review of model compression is provided in the survey by Mishra~\etal~\cite{mishra2020survey}. In our study, we leverage the correlations between latent-grid weights and data values to execute latent-grid compression utilizing error-bounded floating-point compressors.

\subsection{Meta-Learning for INRs}

Meta-learning trains neural networks using examples from specific tasks, enhancing their performance on new, similar tasks. This technique enables networks to rapidly adapt to new tasks with limited data. Notable algorithms in this domain include MAML~\cite{finn2017model} and Reptile~\cite{nichol2018first}. They utilize gradient-based optimization to identify initial weights, optimizing the network's ability to learn new task instances at test time more effectively.  For implicit neural representations, Sitzmann~\etal~\cite{sitzmann2020metasdf} adopted meta-learning to expedite  the learning of new deep signed distance fields. Tancik~\etal~\cite{tancik2021learned} further extended this approach to a broader range of signal types. Our study also prioritizes quick convergence, given the limited time budget for \insitu neural network training. By leveraging the temporal coherence between adjacent data timesteps, we initialize the neural network with weights learned from previous iterations. We refer to this strategy as weight caching.

\subsection{\InSitu Visualization}

The field of \insitu visualization has seen a growing interest in automating the identification of crucial regions for analysis, visualization, and storage. 
To achieve this, researchers use ``\insitu triggers'', defined as Boolean indicator functions, to characterize data features~\cite{bennett2012combining}. 
Larsen~\etal\cite{InSituTrigger} made a significant contribution by introducing the first general-purpose interface for creating \insitu triggers in the Ascent visualization and analysis infrastructure~\cite{Ascent}, simplifying the development process.
The \diva framework~\cite{wu2020diva} enhances the usability of \insitu triggers by enabling reactive programming. It can automatically generate fine-grained \insitu triggers and optimize workflow performance based on user-specified data dependencies and high-level constraints. 
However, reactive programming also assumes the ability to freely retrieve the historical value of any data streams, which is unfeasible for large-scale simulations with prohibitively large volume fields.

Beyond reactive programming, methods to cache temporal data efficiently have been studied separately. Hardware solutions like non-volatile RAMs and burst buffers have been employed~\cite{demarle2021situ} in tools like Paraview~\cite{ParaView} using a sliding buffer technique. However, the success of such hardware-based methods depends heavily on the hardware availability. Software alternatives, like volume data compression algorithms, exist but commonly require decompression to access the data. Our approach uses implicit neural representation to eliminate the need for decompression, offering a more optimal solution.

\begin{figure}[tb]
    \centering
    \includegraphics[width=\linewidth]{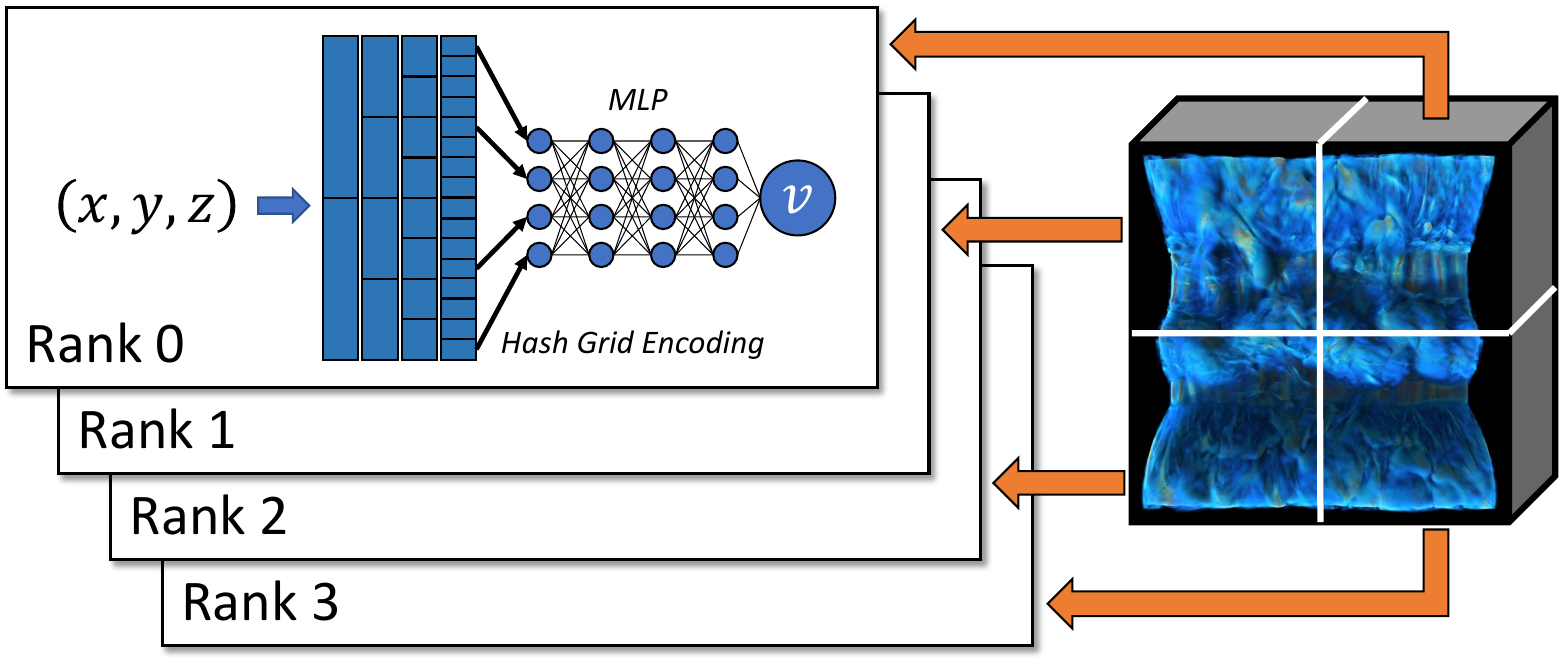}
    \vspace{-2em}
    \caption{\label{fig:network-arch}The design of distributed volumetric neural representation (DVNR).}
    \vspace{-1em}
\end{figure}

\section{Distributed Neural Representation (DNR)}

Implicit neural representation (INR) takes a spatial position $(x,y,z)$ as input and outputs a value $\textbf{v}$ corresponding to the volume field value at that position, as represented by the following equation:
\begin{equation}
\Phi: \mathbb{R}^3 \rightarrow \mathbb{R}^D,~ (x,y,z) \mapsto \Phi(x,y,z) = \textbf{v}.
\end{equation}
Here, $\textbf{v}$ can be a scalar ($D=1$), 3-dimensional vector ($D=3$), or high-dimensional vector ($D>3$), depending on the specific problem. \tvcg{Our base INR architecture comprises a multi-resolution hash encoding layer introduced by M\"{u}ller~\etal~\cite{muller2022instant}} and a small multilayer perceptron (MLP) network. The MLP network employs ReLU activation functions between layers.

The multi-resolution hash encoding layer can encode spatial information efficiently at different resolutions, which is critical for capturing fine details in volume data. Given that the method is a variant of grid-based position encoding, \tvcg{this encoding layer is also referred to as latent-grids.}

The neural network is trained by sampling input coordinates uniformly within the volume domain and computing the corresponding target values through appropriate interpolation methods utilizing reference data. To enhance network stability and accuracy, we normalize both input coordinates and output values to the range of $[0,1]$. This process adjusts the data scale to a consistent range, facilitating more effective learning by the implicit neural network.

Through inference, the neural network can output $\textbf{v}$ on-demand for any arbitrary coordinates within the volume domain, with a low memory footprint. It is also possible to decode the neural representation back to its original grid-based representation, making the technique compatible with existing visualization and analysis infrastructures. For legacy systems that rely on traditional grid-based data structures for visualization and analysis, this can be particularly crucial.

Existing methods on INR can only handle volume data that fully resides on a single machine. In the subsequent sections, we will demonstrate how this technique is adapted for use in distributed scenarios.

\subsection{Supporting Distributed Data via Model Parallelism}

We have developed a decentralized methodology to construct INRs for distributed volume data. \tvcg{Our methodology operates on the assumption that each of the local data partition on each MPI rank encompasses a box region.} We create and train a base INR model on each rank using the corresponding data partition.
This approach results in numerous INR models being independently generated and distributed across multiple ranks.
To perform data analysis and visualization, we leverage standard parallel computing techniques and the sort-last visualization pipeline~\cite{291528}. To maximize the use of the input range of each local INR, global coordinates are first transformed into relative coordinates of a data partition, and then normalized to the range of $[0,1]^3$. Similarly, we normalize output values to $[0,1]$ using the maximum and minimum values of each local partition to maximize the utilization of INR's output value range. Our approach avoids the need for extra interprocess communications between ranks during the training process, thus leading to good scalability. An overview of our technique is provided in \Cref{fig:network-arch}.

\subsection{Improving Scalability via Adaptive Parameters}

We follow the standard procedure to optimize DVNR. We initiate the process by normalizing the dataset according to its minimum and maximum values. Then, for each training iteration, we first perform stochastic uniform sampling of $N_{\text{batch}}$ 3D coordinates, \tvcg{where $N_{\text{batch}}$ stands for the batch size}. 
\tvcg{We use trilinear interpolation to get the data value at an arbitrary coordinate within the volume domain.} 
Next we compute the corresponding data values for each sample position using the ground truth volume data. Finally we use the sampled coordinate-value pairs to optimize the neural network. The batch size $N_{\text{batch}}$ is adjustable by the users.

Since the ``training set'' derived from the continuous 3D domain is theoretically infinite, we manually define the number of training iterations $N_{\text{train}}$ required. We linearly adjust $N_{\text{train}}$ in proportion to the total voxel count $N_{\text{vox}}$ in each local partition, ensuring scalable training across different data sizes. 

However, ending the training process solely based on a fixed iteration count can sometimes lead to inefficiencies. To address this, our approach also allows for the training to be terminated when the moving average of the loss value falls below a predetermined threshold set by the user.

Additionally, users can specify a minimum number of training iterations $N_{\text{train}}^{\text{min}}$ to ensure adequate compression quality for smaller volume datasets. Therefore, the exact formula to calculate the maximum number of training iterations is $N_{\text{train}}^{\text{max}} = \text{max}(N_{\text{train}}^{\text{min}}, \lceil N_{\text{vox}} / N_{\text{batch}} \rceil \cdot N_{\text{epoch}})$, \tvcg{where $N_{\text{epoch}}$ is a parameter providing additional control of the training process.}

\tvcg{Additionally, we scale latent-grid resolutions and the hash table size $T$ based on the local and the global voxel counts, $N_{\text{vox}}$ and $N_{\text{vox}}^{\text{global}}$ respectively}. This adjustment is particularly crucial in addressing strong scaling problems: as we increase the total number of MPI ranks, the overall model size would also increase proportionally, leading to a linear decrease in compression ratio. In our INRs, most of the network weights are stored in latent-grids. By proportionally adjusting grid resolutions, we maintain a relatively consistent compression ratio across varying data sizes. 
In our implementation, \tvcg{we also define a minimum hash table size $T_{\text{min}}$} to avoid model collapsing during training. \tvcg{The hash table size $T$} is scaled according to the formula $T=\text{max}(T_{\text{min}}, T_{\text{ref}} \lceil N_{\text{vox}} / N_{\text{vox}}^{\text{global}} \rceil)$, \tvcg{where $T_{\text{ref}}$ stands for a reference hash table size controlled by the user}. Then, we scale all the latent-grid resolutions using the base resolution $R_{0}=\lfloor  R_{\text{ref}} \sqrt[3]{T / T_{\text{ref}}} \rfloor$, \tvcg{where $R_{\text{ref}}$ is a user-defined scaling factor}.

\begin{figure}[tb]
  \centering
  \includegraphics[width=0.9\linewidth]{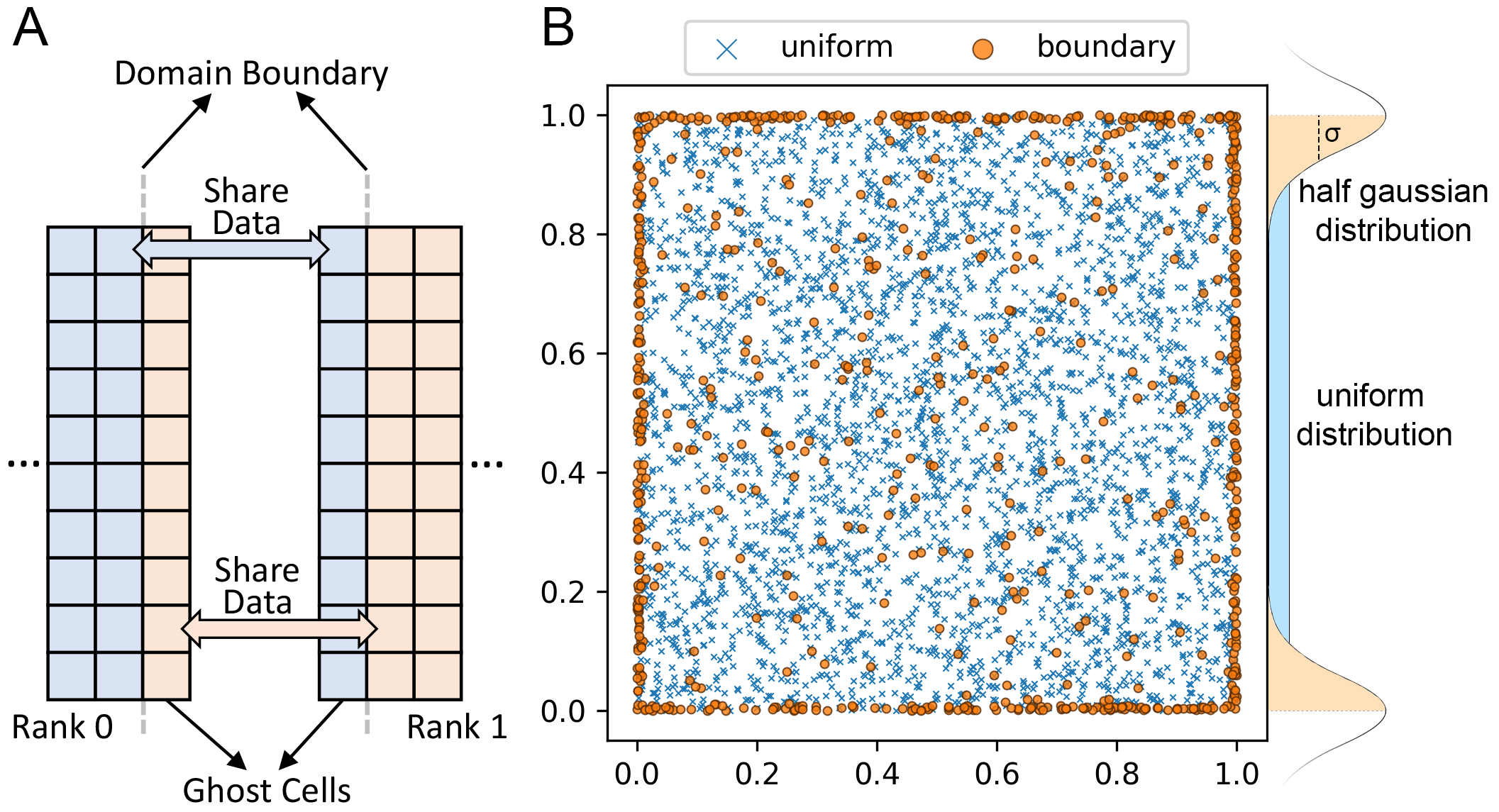}
  \vspace{-1em}
  \caption{\label{fig:sample-distribution}A) Illustration of ghost cells \tvcg{used in} DVNR training \tvcg{(assuming cell-centered volume discretizations~\cite{diskin2011comparison} in the illumination)}. B) In DVNR training, boundary connectivity is enhanced by utilizing training samples from two distinct distributions: uniform and boundary-centered half-Gaussian. More specifically, $L1$ losses are computed for samples from each distribution, which are then combined through a weighted average.}
  \vspace{-1em}
\end{figure}

\subsection{Enforcing Boundary Connectivity via Boundary Loss}

To maintain the continuity of the neural representation across partition boundaries during training, we employ two main techniques. Firstly, we incorporate a single layer of ghost cells into the training process. Ghost cells are one or more layers of grid cells surrounding the external boundary of the local partition, as highlighted in \Cref{fig:sample-distribution}A. Owned by one process but replicated across adjacent ones, they provide accurate boundary information for each local data partition.
Notably, these cells are typically precomputed in data-distributed simulations and visualization infrastructures, thus eliminating the need for extra interprocess communications.

\begin{figure}[tb]
  \centering
  \includegraphics[width=\linewidth]{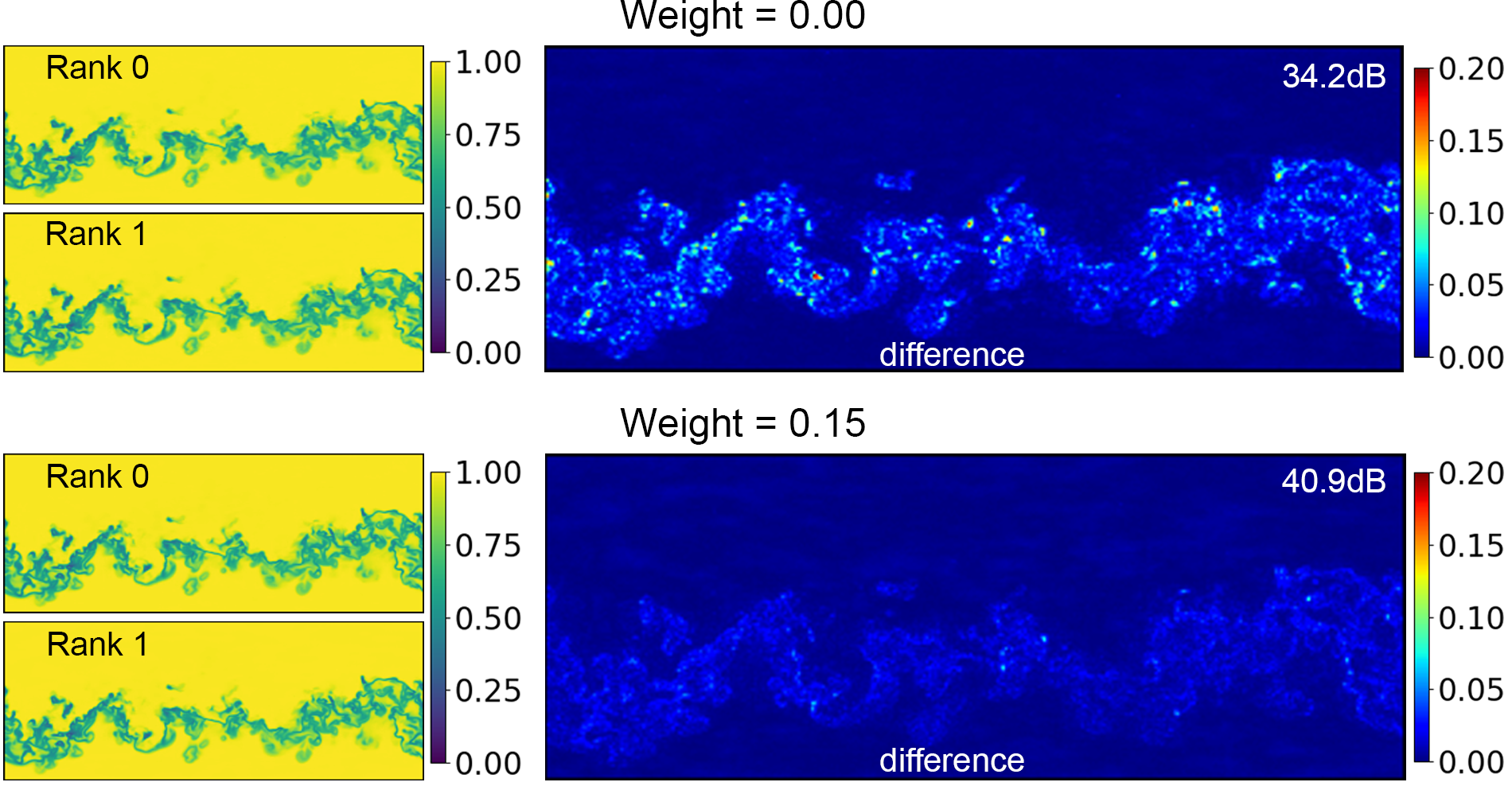}
  \vspace{-2em}
  \caption{\label{fig:boundary-comparison}This figure illustrates a comparison between boundary slices of two adjacent partitions using different weighting factors (\tvcg{referred to as} $\lambda$). It features pseudo-color plots that highlight the differences between the slices. Both neural representations were trained for 10k steps using the S3D heat release field.}
  \vspace{-1em}
\end{figure}

Secondly, an additional $L1$ loss that prioritizes the continuity of the neural representation on partition boundaries is incorporated, referred to as the boundary loss. The boundary loss is calculated by drawing samples from half Gaussian distributions centered on volume domain boundaries, as shown in \Cref{fig:sample-distribution}B. The sampling probability density function is:
\begin{equation}
P(X) = \frac{1}{C}\sum_{i = x,y,z} \text{exp}(-\frac{X_i^2}{2\sigma^2}) + \text{exp}(-\frac{(X_i-1)^2}{2\sigma^2}),
\end{equation}
where $X_i \in [0,1]$ is one of the three components from the sampled coordinate $X$, $\sigma$ is the standard deviation that controls the spread of samples, and $C = {6\sigma\sqrt{2\pi}}$ is the normalization factor.
Consequently, the overall loss $L$ is a weighted sum:
\begin{equation}
L = (1 - \lambda) ~ L_{1}(X_{\text{Uni}}, Y_{\text{Uni}}) + \lambda ~ L_{1}(X_{\text{Bound}}, Y_{\text{Bound}}),
\end{equation}
where $X_{\text{Uni}}$ and $Y_{\text{Uni}}$ are the reference and predicted volume values obtained from uniform samples, and $X_{\text{Bound}}$ and $Y_{\text{Bound}}$ are values from aforementioned boundary samples. $\lambda$ is a weighting factor that controls the influence of the boundary term on the overall loss. In \Cref{fig:boundary-comparison}, we highlight the effect of boundary loss on training accuracy. We calculate the total loss $L$ by first drawing $(1 - \lambda)N_{batch}$ uniform samples and $\lambda N_{batch}$ boundary samples. Then we perform data interpolation using these samples. Finally we compute $L$ as a standard unweighted $L1$ loss. This modification ensures that DVNR consistently draw $N_{batch}$ training samples per iteration, rendering the training cost unaffected by the value of $\lambda$.
\tvcg{We determine the value of $\lambda$ and $\sigma$ through a parameter study in \Cref{dvnr:sec:ablation_boundary_loss}.}

\subsection{Enhancing Compression Ratio via Model Compression}

\begin{figure}[tb]
  \centering
  \includegraphics[width=\linewidth]{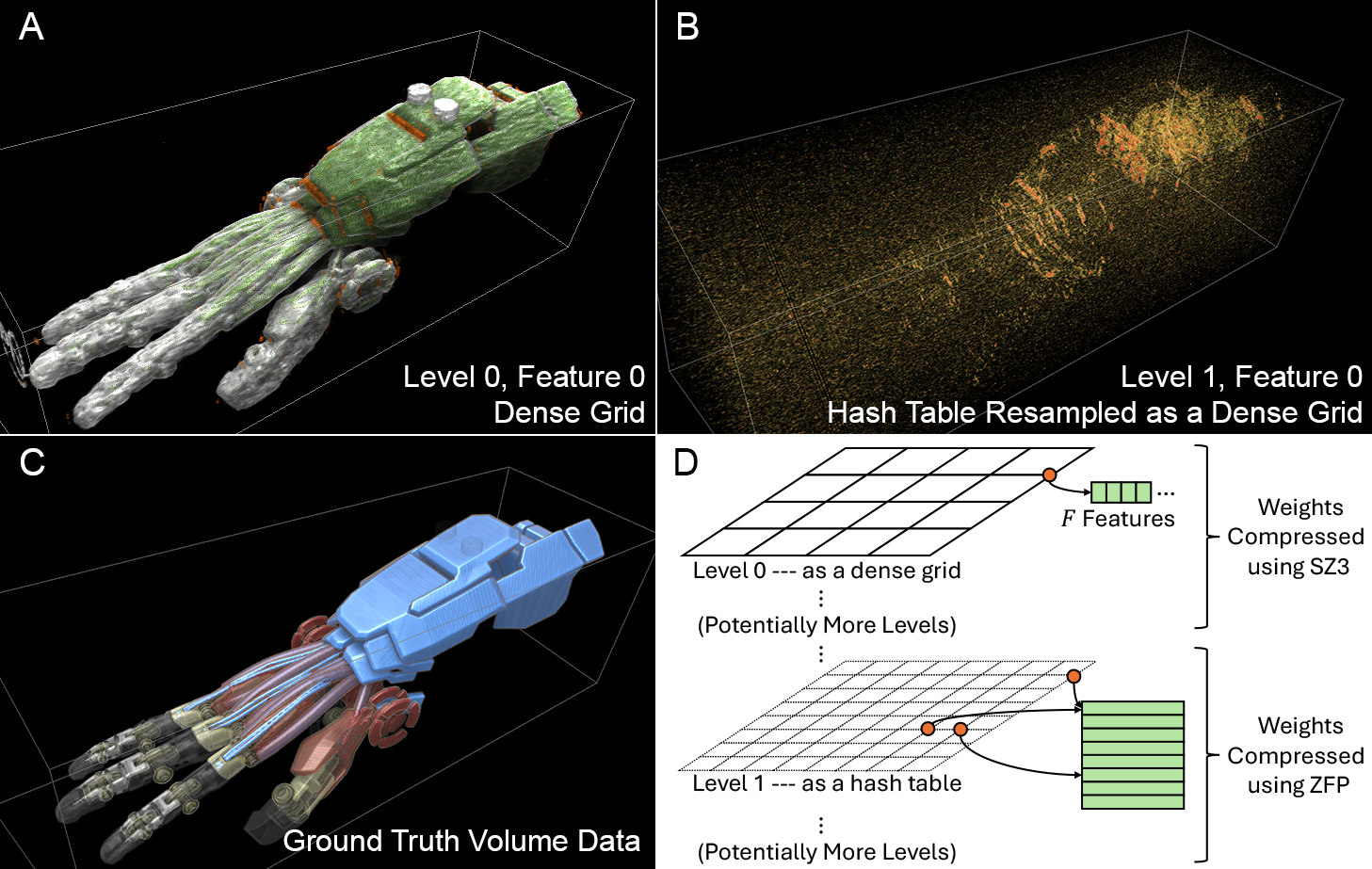}
  \vspace{-2em}
  \caption{\label{fig:hash21}This figure presents volume visualizations of multi-resolution hash encoding weights arranged into 3D grids. A) The visualization at a dense resolution clearly reveals the shape of the underlying object within the volume data. B) At a hashed resolution, the visualization shows some noise, yet the fundamental structure of the object is still recognizable. C) For reference, the volume visualization of the actual data is displayed. D) The diagram outlines our model compression strategy, applying 3D \szthree compression~\cite{liang2022sz3} to dense resolutions and 1D \zfp compression to hashed resolutions. In this example, the \textbf{Mechanical Hand} dataset is compressed via a single INR that employs a two-level multi-resolution hash encoding, yielding a 40.7dB PSNR \tvcg{reconstruction quality after compression}.}
  \vspace{-1em}
\end{figure}

In addition to training the neural network, we also perform \tvcg{additional compressions on model weights, referred to as \emph{model compression}, to further improve the overall compression ratio.} As pointed out by Wu~\etal~\cite{wu2023hyperinr}, for grid-encoding-based INRs, there is a strong correlation between encoder weights and data values. We illustrate this in \Cref{fig:hash21}, where direct volume renderings of trained encoder weights and the ground truth data values are compared. For dense grid levels where encoder weights are stored using single 3D arrays, volume structures visible in the ground truth rendering can also be easily identified in the renderings of weights. For hash grid levels where weights are stored using hash tables, \tvcg{we resampled the weights to a dense grid of the corresponding resolution for visualization and found that}
it is however very difficult to classify corresponding volume structures. This observation motivates us to develope a model compression method utilizing traditional floating-point lossy compressors.

Specifically, three types of compressors are involved in our technique. We reinterpret dense encoder weights as 4D arrays of size $R\times R\times R\times F$ where $R$ is the grid resolution and $F$ is the number of features per grid level, and compress them using the \szthree compressor~\cite{liang2022sz3} \tvcg{targeting accuracy $r_1$.} We reinterpret hash encoder weights as 2D arrays of size $T\times F$ with $T$ being the hash table length, and compress them using the \zfp compressor~\cite{lindstrom2006fast, lindstrom2014fixed} \tvcg{targeting accuracy $r_2$.} We also compress all the MLP weights as an 1D array using the \zfp compressor, \tvcg{targeting accuracy $r_3$.} Finally, we merge all the compressed byte streams and compress them using  \zstd~\cite{Zstandard}. \tvcg{The accuracy targets $r_1,r_2,r_3$ are directly controlled by users.} Since \zfp, \szthree and \zstd are all highly optimized compressors, and they are applied to neural network weights data which are already relatively small, the added model compression overheads are expected to be small, making it suitable for \insitu visualization. Notably, in our implementation, model weights are stored as 16-bit floats. \tvcg{Due to the lack of support for 16-bit floats in \zfp and \szthree, we currently promote model weights to 32-bit floats before compressing them in our current implementation.} These model weights are substantially smaller than the volume data, often on the order of kilobytes, making the memory overheads of data conversion also negligible. \tvcg{The model compression ratios are computed by comparing the size of the unpromoted 16-bit model with the compressed bytestream.}

During our development, we also experimented using \tthresh~\cite{ballester2019tthresh} for \tvcg{model compression.} However, \tvcg{\tthresh cannot efficiently compress latent-grids weights, because typical latent-grids are often too small for \tthresh to amortize the cost of storing factor matrices.}

\subsection{Optimizing Training Time via Weight Caching}

We also apply learned weight initialization to accelerate DVNR training. In \insitu visualization, we typically apply a data compression algorithm to the same set of fields for all the simulated timesteps. Although data from different timesteps are different, the actual difference between adjacent timesteps are usually small; therefore, we can leverage the neural network weights learned from a previous timestep to initialize the network for the current one.

In our method, such a weight initialization is implemented as a temporal cache. Entries in the cache are distinguished based on the name of the volume field being compressed as well as the neural network configuration.
After a DVNR compression is performed, learned neural network weights will be recorded in the corresponding cache entry. In the next iteration, the newly constructed neural network is initialized using these cached weights.

\subsection{Implementation}

Our DVNR implementation consists of a base INR model and several shared modules for tasks including model compression, weight caching, data sampling, and I/O. These shared modules are all implemented in C++ and accelerated via CUDA if possible. The base INR model is implemented in both Python and C++, and leverages GPU tensorcores to accelerate the model training process through \tvcg{the} \tcnn~\cite{tcnn} library. The C++ backend provides better performance but currently only implements one INR architecture. The Python backend enables quick prototyping and testing of new INR designs, which is implemented by embeding the Python interpreter into the C++ host program via 
the pybind11 library.
\tvcg{In our implementation}, we use the C++ \tvcg{backend} unless specified otherwise.

Users can directly control the base INR configuration, and we utilize the Adam optimizer with exponential learning rate decay to train the model. We set $\beta_1=0.9$, $\beta_2=0.999$, $\epsilon=1E-8$, and $L2$ weight decay $=1E-9$. We also allow users to specify the initial learning rate and hyperparameters related to learning rate decay. For the loss function, our default weighting factor is $\lambda=0.15$ and the standard deviation for boundary sampling is $\sigma=0.005$. We discuss the choice of these parameters in~\Cref{dvnr:sec:ablation}. \tvcg{For model compression, we empirically set $r_\text{enc} = r_1 = r_2$ and $ r_\text{mlp} = r_3$, and then allow users to only control $r_\text{enc}$ and $r_\text{mlp}$.} 

\tvcg{We acknowledge that DVNR still requires users to control a few parameters to achieve optimal performance. To address this, we offer several pre-defined settings for users to choose from. While these default settings provide a good starting point, they may not always yield the best performance for DVNR across all datasets. 
To ensure the experiments in this work are as effective as possible, we conducted a small grid search to identify the best parameters for each dataset and compression ratio. The results of this parameter search are provided in the supplementary material.
Furthermore, we recognize the importance of simplifying this process for users and see the development of an automated parameter selection mechanism as a valuable area for future work. This would not only enhance user experience but also improve the robustness and adaptability of DVNR across diverse datasets.}

\section{\diva and Ascent Integration}

Reactive programming provides unrestricted access to the history of time-varying values, posing a significant challenge for some applications due to potential impractical data storage requirements~\cite{Fran}. Addressing this data storage issue represents one of the multiple motivations behind the development of DVNR. To specifically tackle this challenge, we have integrated DVNR into the reactive \insitu programming interface, \diva~\cite{wu2020diva}. Furthermore, we have connected our implementation with the lightweight \insitu visualization and analysis infrastructure, Ascent~\cite{Ascent}, to enable and evaluate its use in real-world \insitu applications. This section elaborates on the details of the integration.

\subsection{Creating and Accessing DVNR}

We developed a specialized reactive constructor in \diva to encapsulate normal volume fields and create DVNR models. This process kick-starts the DNR training  in C++ or PyTorch, which continues until the user-defined stopping criteria are met. After training concludes, access to the original volume field is revoked, and the trained neural network is transferred and stored in system RAM for future use.

As previously mentioned, our DVNR training technique is free from interprocess communications, and we normalize network input and output ranges to $[0,1]$, while recording the original value ranges for reference. The input ranges, recomputable from the partition geometry, don't require explicit storage. However, for each DVNR-compressed field, storing the output value range is crucial, as it is useful for adjusting transfer function values and ensuring accurate visualizations. Notably, the DVNR training process is referentially transparent, albeit with potential training variances. This property enables full utilization of \diva's lazy evaluation, allowing for the automatic bypassing of DVNR construction if not accessed by any triggers \tvcg{from any} ranks.

DVNR can train on volume domains of arbitrary shapes. However, our sort-last visualization pipeline currently does not support the composition of image tiles from non-rectangular partitions. Therefore, we  confine our DVNR application to simulations with rectangular domain decompositions. Such a constraint is also present in other state-of-the-art parallel rendering libraries like OSPRay~\cite{wald2016ospray}. We plan to support other domain decomposition schemes later.

To prevent data duplication, we use zero-copy arrays wherever possible. We also developed callback-based training data samplers for both scalar and vector fields, and for all volume mesh types used in Ascent. For structured meshes, we transfer the data to the GPU and generate training samples using customized CUDA interpolation kernels. For unstructured meshes, we leverage parallel primitives from VTKm~\cite{VTKm} to access the data. Sampling from a structured mesh is more efficient, so we also offer the option to resample an unstructured mesh into a structured one. Optimizing our unstructured data sampler is a topic for future work. In this work, we enable the resampling process for unstructured volume data. This benefits not only DVNR but also other compressors.

We infer DVNR models to access compressed volume values. Algorithms optimized specifically for DVNR can perform these network inferences on-demand, reducing the overall memory footprint. However, to maintain compatibility with existing visualization algorithms, we also provide the option to decode a DVNR model back to its original grid form. \diva's runtime automatically selects the appropriate inference method. In the following subsections, we highlight existing operations optimized specifically for DVNR. We plan to introduce DNR-compatible routines for more visualization algorithms in the future.

\subsection{Achieving Temporal Caching via DVNR}\label{sec:temporal_caching}

We implement temporal caching by modifying the sliding window operations in \diva. This operator takes an input volume field and an integer size, transforming the time-varying volume field into a temporal array. Users can use it like a normal array and perform array programming for visualization and analysis. Internally, for each timestep, the sliding window operator records newly trained DVNR models by storing them in a static C++ STL vector. When the number of cached models reaches the user-defined threshold, the oldest model is removed to accommodate new entries. Compared to the prior approach of caching volume fields as uncompressed 3D grids, utilizing DVNR not only avoids excessive memory consumption but also enables the creation of substantially larger sliding windows.

\subsection{Distributed Volume Visualization}

Direct volume rendering and isosurface extraction are two key operations in volume visualization. Therefore, we have developed implementations specifically optimized for these processes, which can directly utilize DVNR models, thereby eliminating the need for decoding. This section delves into the specifics of our implementations.

For direct volume rendering, we utilize the sample-streaming method developed by Wu~\etal~\cite{wu2022instant} as the \tvcg{base rendering algorithm.}
\tvcg{This method divides the ray marching algorithm into smaller GPU kernels, handling coordinate generation, model inference, and shading through separate kernel launches. This approach enhances parallelism in the model inference step, thereby improving overall rendering performance.} 
Building upon this method, we further developed a sort-last parallel rendering pipeline to effectively accommodate distributed neural representations. Our approach avoids reverting the DVNR model back to a grid, thereby maintaining a minimal memory footprint. However, it is important to note that neural network inference is inherently more computation-intensive than accessing a 3D grid, which leads to higher rendering latency in our renderer.

We also offer an isosurface extraction routine that is fully compatible with DVNR models. This routine employs a customized CUDA inference kernel, akin to those reported by Weiss~\etal~\cite{weiss2021fast} and Wu~\etal~\cite{wu2022instant}, facilitating  the on-demand retrieval of volume values from a DVNR model, thereby minimizing the memory footprint. The extraction process is executed locally on each MPI rank, explicitly generating triangle meshes. These meshes  are then zero-copy transferred to Ascent for distributed rendering. The subsequent subsection details the data transfer process as well as the interoperation between DVNR and Ascent.

\begin{figure}[tb]
    \centering
    \includegraphics[width=0.8\linewidth]{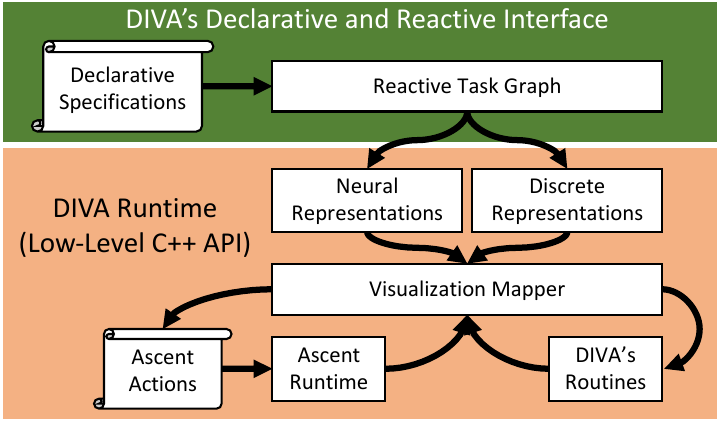}
    \vspace{-1em}
    \caption{\label{fig:diva-ascent}The overview of our DIVA-Ascent integration.}
    \vspace{-1em}
\end{figure}

\subsection{DVNR and Ascent Interoperability}

Ascent is a widely adopted, lightweight \insitu visualization and analysis infrastructure~\cite{Ascent}. To enable the integration of DVNR in real-world simulations, we have also incorporated Ascent into our system. \Cref{fig:diva-ascent} provides an overview of our design. Our integration features  a bidirectional in-memory communication mechanism between \diva and Ascent, which facilitates the expression of key Ascent concepts as \diva operators. It also allows the \diva runtime to dynamically generate Ascent zero-copy actions for executing various operations. This synergy makes smooth interoperability between DVNR and Ascent possible. \tvcg{Visualization results generated by our integration are showcased in the supplementary material}.

\section{Evaluation}

This section presents an evaluation of DVNR, focusing on its performance, scalability, quality, memory footprint, and usability. Our analysis utilized eight \posthoc volume datasets. We also applied DVNR to three distinct \insitu simulations to demonstrate its versatility: \textbf{Cloverleaf}, a simulation that solves the compressible Euler equations on a Cartesian grid~\cite{mallinson2013cloverleaf}; \textbf{NekRS}, a GPU-ccelerated spectral element Navier-Stokes Solver for incompressible turbulent flows employing an unstructured hexahedral mesh~\cite{fischer2022nekrs}; and \textbf{S3D}, a scalable direct numerical solver for reactive and compressible flows, based on a rectilinear mesh~\cite{hawkes2007scalar, chen2009terascale}.

Our evaluations were conducted on the Polaris supercomputer, located at the Argonne Leadership Computing Facility. Each node of Polaris features 4 NVIDIA A100 GPUs, an AMD EPYC Milan 32-Core CPU, and 512GB of system RAM. We configured all experiments to run with 4 MPI ranks per node, dedicating 1 GPU and 8 CPU cores to each rank.

\subsection{Scalability}

\begin{figure}[tb]
  \centering
  \includegraphics[width=\linewidth]{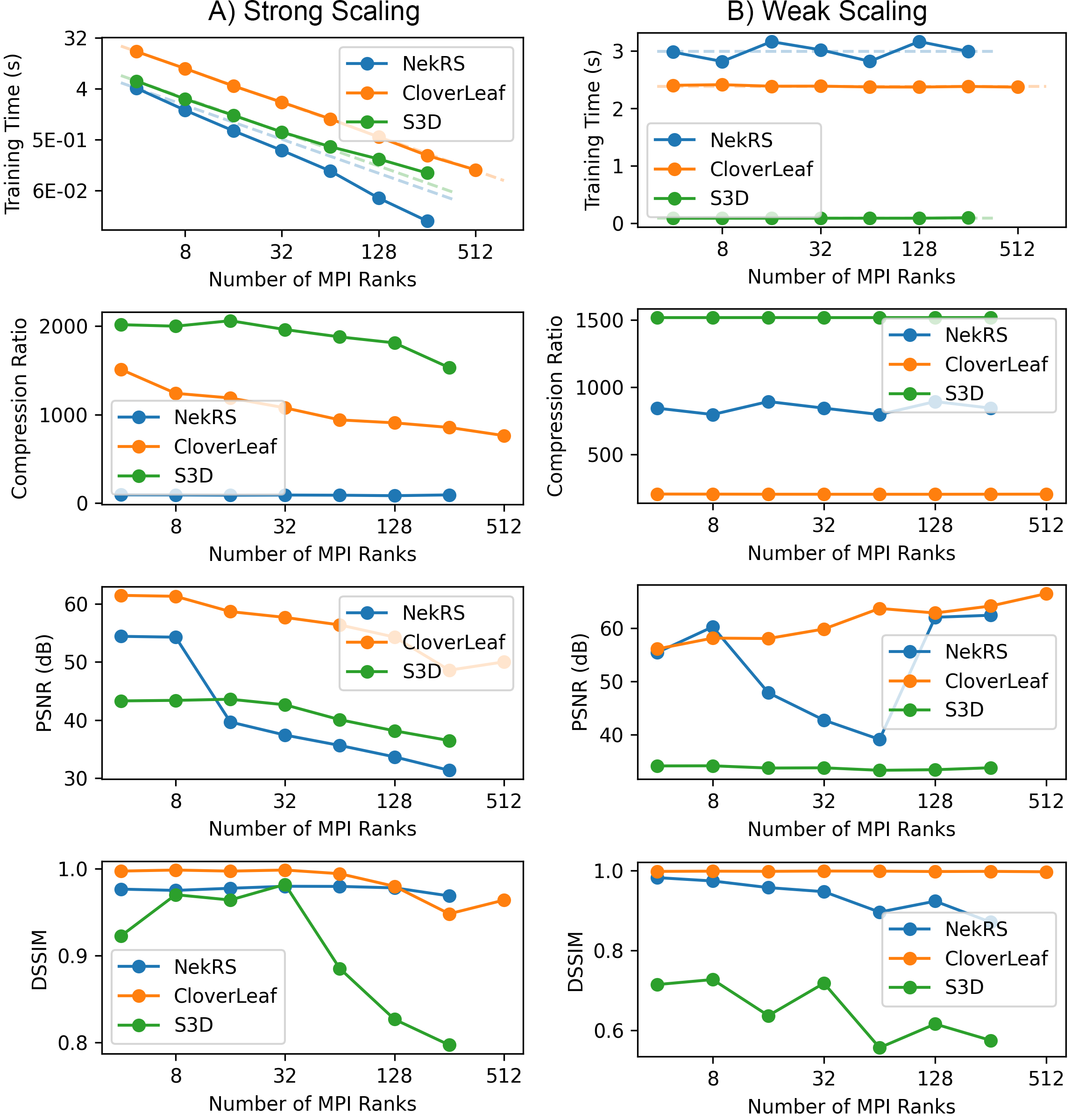}
  \vspace{-2em}
  \caption{\label{fig:scaling}Strong and weak scaling results. For strong scaling, the simulation domain sizes were $1000^3$ (CloverLeaf), $400^3$ (NekRS), and $640^3$ (S3D). For weak scaling, the maximum domain sizes were $2580\times2580\times2580$ (CloverLeaf), $1680\times1680\times1680$ (NekRS), and $10368\times320\times3840$ (S3D). More specifically, the per rank data resolution for CloverLeaf and S3D were $512\times256\times256$ and $324\times320\times240$ respectively. For NekRS, the resolution fluctuated around $260^3$ because NekRS requires cubic domains. \tvcg{Due to space limitations, we present only PSNR and DSSIM measurements. SSIM measurements are available in the supplementary material.}}
  \vspace{-1em}
\end{figure}

Our scalability study of DVNR involved both strong and weak scaling tests applied to all three simulations, scaling them up to 128 nodes (\ie 512 MPI ranks). Each simulation ran for a set number of timesteps at the maximum domain resolution supported by the available system RAM. During each timestep, DVNR compressed one of the volume fields. We controlled the total number of training steps across all timesteps for consistency. To isolate the scalability of the network training process, we excluded weight caching from these tests. Our measurements included the average DVNR compression time, \tvcg{Peak Signal-to-Noise Ratio (PSNR), Structural Similarity Index Measure (SSIM)~\cite{wang2004image}, Data SSIM (DSSIM)~\cite{baker2022dssim},} and compression ratio, aggregated across all timesteps. These results are detailed in \Cref{fig:scaling}.

\subsubsection{Training time}
In strong scaling tests, DVNR demonstrated excellent scalability across all three simulations. Notably, DVNR's training times often fell below the ideal scaling curve when scaling across a large number of MPI ranks. This occurred because, to maintain a relatively consistent compression ratio in strong scaling, we proportionally decreased the network size as the simulation scaled up, which consequently sped up backpropagation passes. In weak scaling tests, DVNR also exhibited strong scalability. However, training times for the NekRS case occasionally deviated from expectations. This deviation is attributed to NekRS's requirement for all local partitions to be perfect cubes, which led to imperfect scaling of the simulation domain due to rounding errors in domain size calculations.

\subsubsection{Reconstruction Quality}
In both strong and weak scaling, DVNR consistently maintained good reconstruction quality in terms of PSNR, \tvcg{SSIM, and DSSIM} across various configurations.
\tvcg{For strong scaling, we observed a slight decline in data reconstruction quality as the simulation scaled up. This is because, as the volume domain was divided into more partitions, we also proportionally scaled down the model size and training time. The net effect was a minor decrease in overall reconstruction quality.}

\tvcg{For weak scaling, we observed a small increase in quality as CloverLeaf scaled up. This improvement was due to the decrease in data complexity per partition as the number of partitions increased, allowing the neural network to converge faster. Conversely, for NekRS, a decrease in quality was observed between 16 to 64 MPI ranks. We believe this decrease was related to the alignment of the neural network boundaries with the data features. The neural network seemed to learn specific NekRS features more effectively when its boundaries aligned perfectly with the data periodicity. However, for S3D, whose data is highly complex throughout, the data reconstruction quality remained consistently steady.}

\subsubsection{Compression Ratio (CR)}
In strong scaling, a downward trend in the compression ratio was observed. This is attributed to the fact that while the latent-grid resolution can be decreased linearly, the MLP size cannot be reduced in a similar manner. Consequently, as the number of ranks increases, the generation of more MLPs leads to a slight increase in the overall model size. \tvcg{In contrast, weak scaling exhibited nearly ideal results in terms of the compression ratio. The NekRS simulation was the only exception, with its compression ratio fluctuating slightly as the number of MPI ranks increased. This fluctuation is primarily linked to the previously mentioned rounding errors in domain size computations.}

\begin{figure*}[tb]
  \centering
  \vspace{-1em}
  \includegraphics[width=\textwidth]{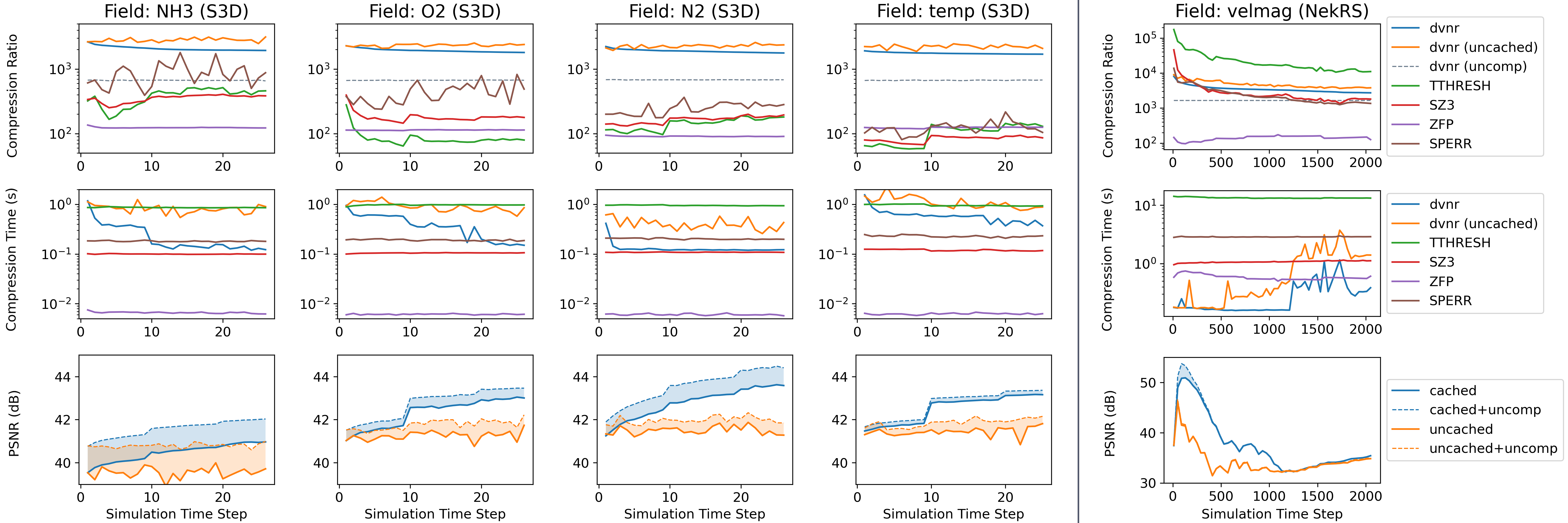}
  \vspace{-2em}
  \caption{In this study, we compared DVNR with compression algorithms \zfp, \szthree, \tthresh and \sperr \insitu. \tvcg{As part of our ablation study (see \Cref{dvnr:sec:ablation}), we also compared DVNR with its variants that disabled the weight caching component (uncached, see \Cref{dvnr:sec:weight_caching}) and those that did not perform model weight compression (uncomp, see \Cref{dvnr:sec:weight_compression}).} We used S3D, configured on a $864\times640\times640$ domain with 64 MPI ranks, compressing at each of its 25 timesteps, and NekRS on a $420 \times 420 \times 420$ domain with 4 MPI ranks, compressing every 10 of its 2000 timesteps. DVNR was targeted to user-defined accuracies of roughly 40dB for S3D and 34dB for NekRS. We aligned \zfp, \szthree, \tthresh, and \sperr to DVNR's PSNR.}
  \label{fig:time_evo}
  \vspace{-1em}
\end{figure*}

\begin{figure*}[tb]
  \hspace*{0.05em}
  \includegraphics[width=0.96\linewidth]{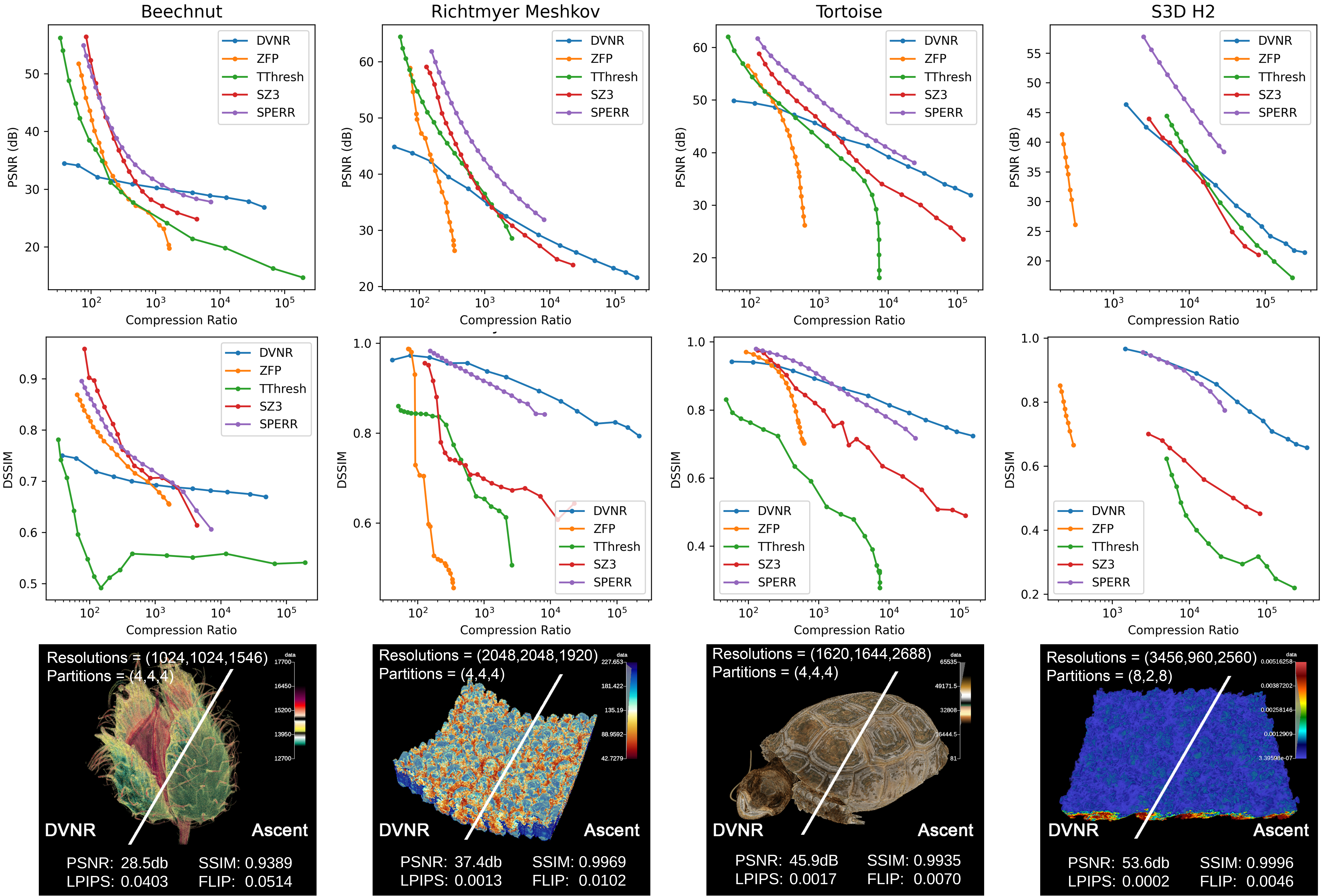}
  \vspace{-0.9em}
  \caption{\Posthoc evaluations using 8 volume datasets with different compression ratios. \tvcg{Compression qualities are reported in terms of volume space PSNR and DSSIM. Additionally, we compare volume rendering results of these datasets using DVNR and Ascent. We measure the the rendering quality in terms of image space PSNR, SSIM, AlexNet-based Perceptual Metric (LPIPS)~\cite{zhang2018perceptual}, and \FLIP~\cite{Andersson2020}. Results for four of the larger datasets are shown in this figure. 
  Remaining results are available in the supplementary material.}}
  \label{fig:static_volume}
  \vspace{-0.5em}
\end{figure*}

\subsection{Comparison Against Traditional Compressors}

\begin{table}[tb]
\vspace{-0.8em}
\setlength{\tabcolsep}{5pt}
\caption{\tvcg{Comparison of data compression time across 
datasets. Measured values are averaged over multiple timesteps for runs detailed in \Cref{fig:time_evo}, 
and averaged over different network configurations for datasets highlighted in \Cref{fig:static_volume}.}}
\label{tab:ratios}
\vspace{-1em}
\scriptsize\centering%
\begin{tabu}{cc rcrrr}
\toprule 
  \multicolumn{2}{c}{} 
& \multicolumn{5}{c}{{
    Average Data Compression Time (seconds)
}} \\
\cmidrule(lr){3-7}
\multicolumn{2}{c}{Dataset} &
\multicolumn{1}{r}{DVNR} & 
\multicolumn{1}{c}{\zfp} & 
\multicolumn{1}{r}{\szthree} & 
\multicolumn{1}{r}{\tthresh} &
\multicolumn{1}{r}{\sperr} \\
\midrule
\multicolumn{2}{r}{Magnetic~\cite{magnetic_reconnection}      } & 1.87   & 0.011 & 0.247 & 1.43   & 0.687 \\
\multicolumn{2}{r}{Rayleigh Taylor~\cite{miranda}             } & 2.98   & 0.039 & 0.545 & 2.97   & 0.657 \\
\multicolumn{2}{r}{Richtmyer Meshkov~\cite{richtmyer_meshkov} } & 13.21  & 0.129 & 3.128 & 15.44  & 0.815 \\
\multicolumn{2}{r}{S3D H2~\cite{s3d}                          } & 5.70   & 0.055 & 0.870 & 5.83   & 1.136 \\
\multicolumn{2}{r}{Pawpawsaurus~\cite{pawpawsaurus}           } & 4.25   & 0.048 & 0.732 & 4.55   & 0.837 \\
\multicolumn{2}{r}{Chameleon~\cite{chamaeleo}                 } & 3.16   & 0.014 & 0.499 & 3.59   & 0.661 \\
\multicolumn{2}{r}{Beechnut~\cite{beechnut}                   } & 2.29   & 0.014 & 0.373 & 2.50   & 0.673 \\
\multicolumn{2}{r}{Tortoise~\cite{tortoise}                   } & 12.38  & 0.066 & 1.550 & 12.23  & 1.112 \\
\midrule  
\multirow{4}{*}{ \begin{tabular}{@{}c@{}}S3D \\ (\insitu)\end{tabular} }
& NH3  & 0.76 & 0.04 & 0.08 & 0.94 & 0.18 \\
& O2   & 0.88 & 0.06 & 0.08 & 0.88 & 0.19 \\
& N2   & 0.34 & 0.07 & 0.09 & 0.91 & 0.20 \\
& Temp & 0.98 & 0.06 & 0.10 & 0.84 & 0.23 \\
\midrule
NekRS (\insitu)
& VelMag & 0.30 & 0.59 & 1.07 & 13.4 & 2.87 \\
\bottomrule
\end{tabu}%
\vspace{-2em}
\end{table}

In this section, we compare DVNR with three widely-used traditional compressors: \zfp, \szthree, \tthresh, \tvcg{and \sperr}. We conducted both \insitu and \posthoc evaluations, assessing their compression ratio, compression time, and reconstruction quality in terms of PSNR, SSIM \tvcg{and DSSIM}. Although \zfp, \szthree, \tthresh, \tvcg{and \sperr} are not inherently designed for distributed data, supports can be easily implemented by independently applying these algorithms to each local partition, which aligns well with DVNR's method. Compression times, SSIMs, \tvcg{and DSSIMs} were averaged over all partitions. PSNRs were calculated based on the average mean squared errors (MSEs) across all partitions. Compression ratios were determined using the global volume size and the cumulative sizes of the compressed outputs. 

\tvcg{All traditional compressors were compiled with OpenMP support and controlled using PSNR targets. Since \zfp does not natively accept PSNR quality controls, the fixed-accuracy mode was used instead, with the ``accuracy tolerance'' iteratively tuned to closely match the PSNR target.~\footnote{\tvcg{We carefully excluded the iterative tuning process from the \zfp compression time measurements. Only the final compression time with the selected ``tolerance'' was recorded.}} To normalize the impact of data dynamic range, input data for all compressors were normalized to the range of $[0, 1]$.}

For the \insitu evaluation, we utilized S3D and NekRS. The simulation configurations were based on recommendations from domain scientists. In each simulation, different fields were compressed using the same DVNR model, with the compression targeted to meet a user-defined accuracy level. Our experimental setups and results are illustrated in \Cref{fig:time_evo}. Additionally, we report the average data compression time for each volume field in \Cref{tab:ratios}.

We found that, with the setups we used, DVNR demonstrated a competitive compression ratio, largely owing to the effective use of model compression techniques. When it comes to compression time, \zfp emerged as the clear leader, \tvcg{followed by \szthree and \sperr.} DVNR was comparable to \tthresh in most cases. However, with the aid of weight caching, DVNR's compression time can be reduced by up to $10\times$ as the simulation evolves.

Our \insitu evaluation was constrained by the specific configurations we chose. To gain a more comprehensive understanding across a broader range of datasets and settings, we also conducted \posthoc evaluations using eight volume datasets, featuring varying compression ratios and quality targets, as illustrated in \Cref{fig:static_volume}. For DVNR, reconstruction quality is influenced by both the model size and the training time. To facilitate a fair comparison with traditional compression models, we varied the model size while controlling the training time to match the average compression times of \tthresh.  The total data compression times for each method are also highlighted in \Cref{tab:ratios}.

\begin{figure*}[tb]
  \centering
  \includegraphics[width=0.93\linewidth]{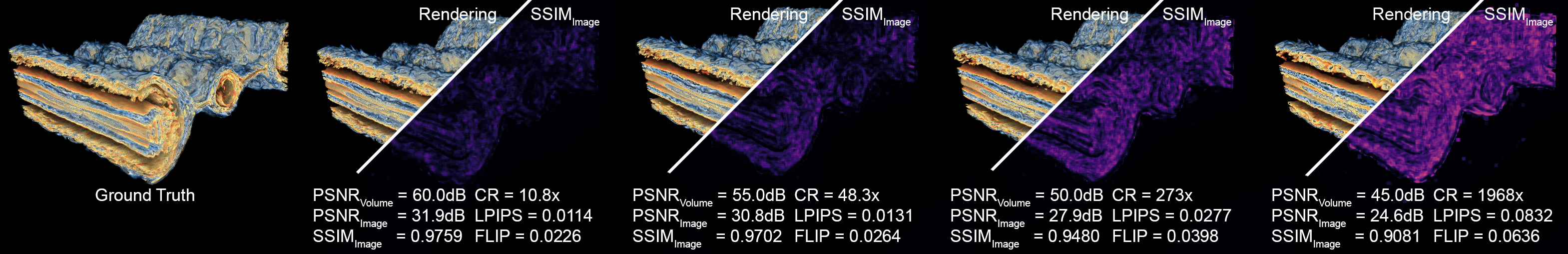}
  \vspace{-0.9em}
  \caption{\tvcg{Visualizations of the \textbf{Magnetic} dataset compressed by DVNR at different PSNR levels. Compression ratios (CR) and various image quality metrics compared against the ground truth rendering are reported. We also compute multiple pixel-wise image qualities, which is available in the supplementary material.}}
  \label{fig:rendering_diffpsnr}
  \vspace{-1em}
\end{figure*}

\tvcg{Our analysis revealed that \szthree, \tthresh, \sperr, and DVNR generally outperformed \zfp in both compression ratio and reconstruction quality. Among them, at higher compression ratios, DVNR consistently delivered better PSNR than \szthree and \tthresh and was the best performer in terms of SSIM and DSSIM.} However, when high reconstruction quality was required, DVNR tended to necessitate a large model, making it less efficient in terms of compression ratio.

While DVNR may not be the fastest compressor nor an absolute winner in compression ratio, it \tvcg{still} offers numerous other advantages. \tvcg{One significant benefit is its capability to provide data access without requiring decompression, which eliminates decompression time compared to traditional compressors and significantly reduces the memory footprint for visualization and analysis tasks.}

\begin{figure}
  \centering
  \vspace{-0.8em}
  \includegraphics[width=0.95\linewidth]{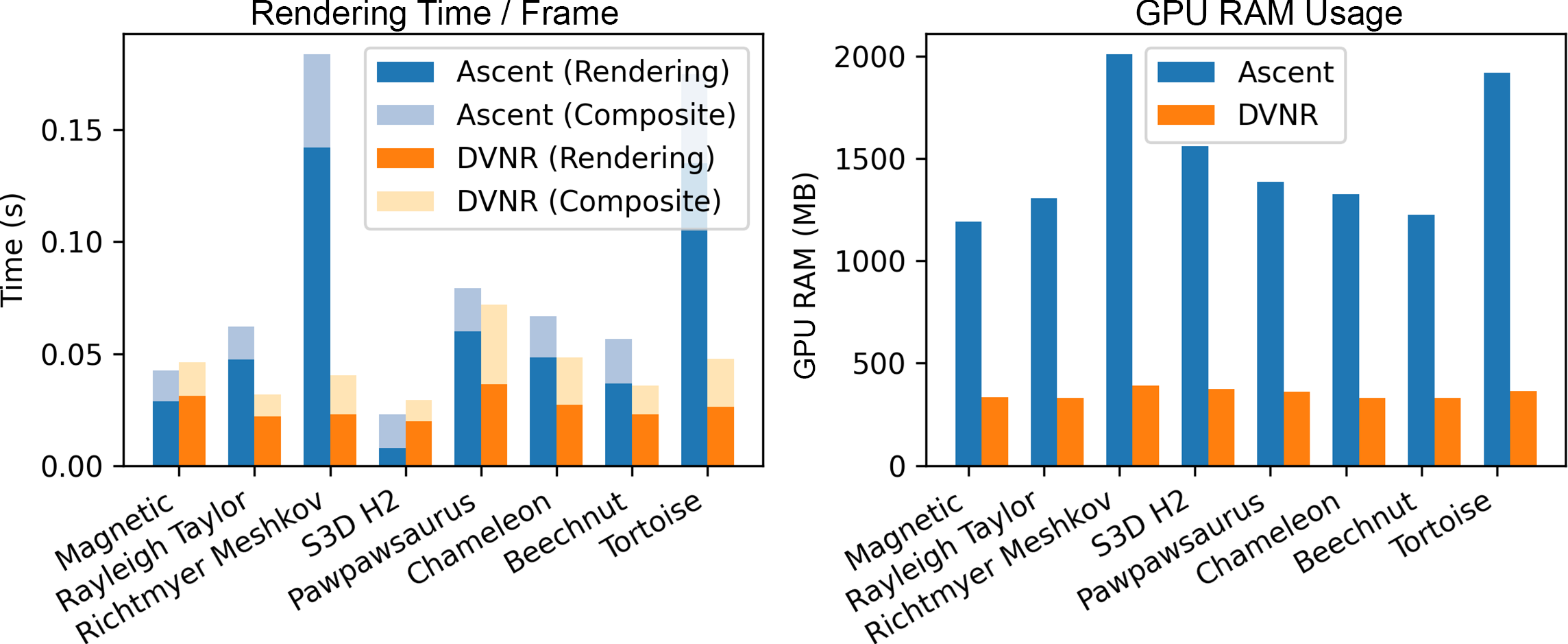}
  \vspace{-0.8em}
  \caption{Profilings of direct volume rendering with DVNR and Ascent. Datasets, configurations and results are shown in \Cref{fig:static_volume}. For fairness, all the data in this experiment shares the same neural network configuration.}
  \label{fig:rendering_statistics}
  \vspace{-1em}
\end{figure}

\subsection{Visualization Comparisons}

This section delves into evaluating the quality of visualizations generated using DVNR.
First, we compared our DVNR-based volume rendering implementation with Ascent's direct volume renderer, specifically utilizing its VTKh backend. The outcomes of these renderings are compared in \Cref{fig:static_volume}. 
\tvcg{Additionally, we compared the visualizations of the same dataset compressed by DVNR at different PSNR levels with results presented in \Cref{fig:rendering_diffpsnr}.}

\tvcgm{We found that DVNR consistently produces high-quality visualization results, with PSNR$_\text{image}$ exceeding 28dB, SSIM$_\text{image}$ above 0.930, LPIPS below 0.040, and  \FLIP under 0.051 compared to the ground truth generated by Ascent.
While the minimal data compression quality required for acceptable visualizations is ultimately dataset-dependent, DVNR generally delivers satisfactory results at volume PSNR levels between 25-30dB. An exception is the \textbf{Magnetic} dataset, which necessitates a PSNR of 40dB.}

Next, to study the rendering performances, we configured the DVNR renderer to directly load pre-trained models. \tvcg{Each testing sequence uses the initial render frame to warm up the system, subsequently followed by the analysis of ten additional frames. We measured the local rendering time averaged across MPI ranks, the global sort-last compositing time, and the peak GPU memory consumption. Results are reported in \Cref{fig:rendering_statistics}.}

We discovered that our DVNR-based rendering implementation matched Ascent's implementation in performance for 5 out of the 8 datasets. In the remaining 3 datasets, our renderer even outperformed Ascent's algorithm. These findings suggest that our DVNR-based algorithm is capable of delivering the necessary performance for \insitu visualization. Regarding \tvcg{GPU memory} consumption, our DVNR-based implementation managed to reduce the GPU memory footprint by up to $80\%$ during rendering. Additionally, it's notable that the GPU memory footprint for volume rendering with DVNR does not increase when handling larger volume data, which is an important advantage for \insitu visualization.

\begin{figure}[tb]
  \centering
  \vspace{-0.4em}
  \includegraphics[width=\linewidth]{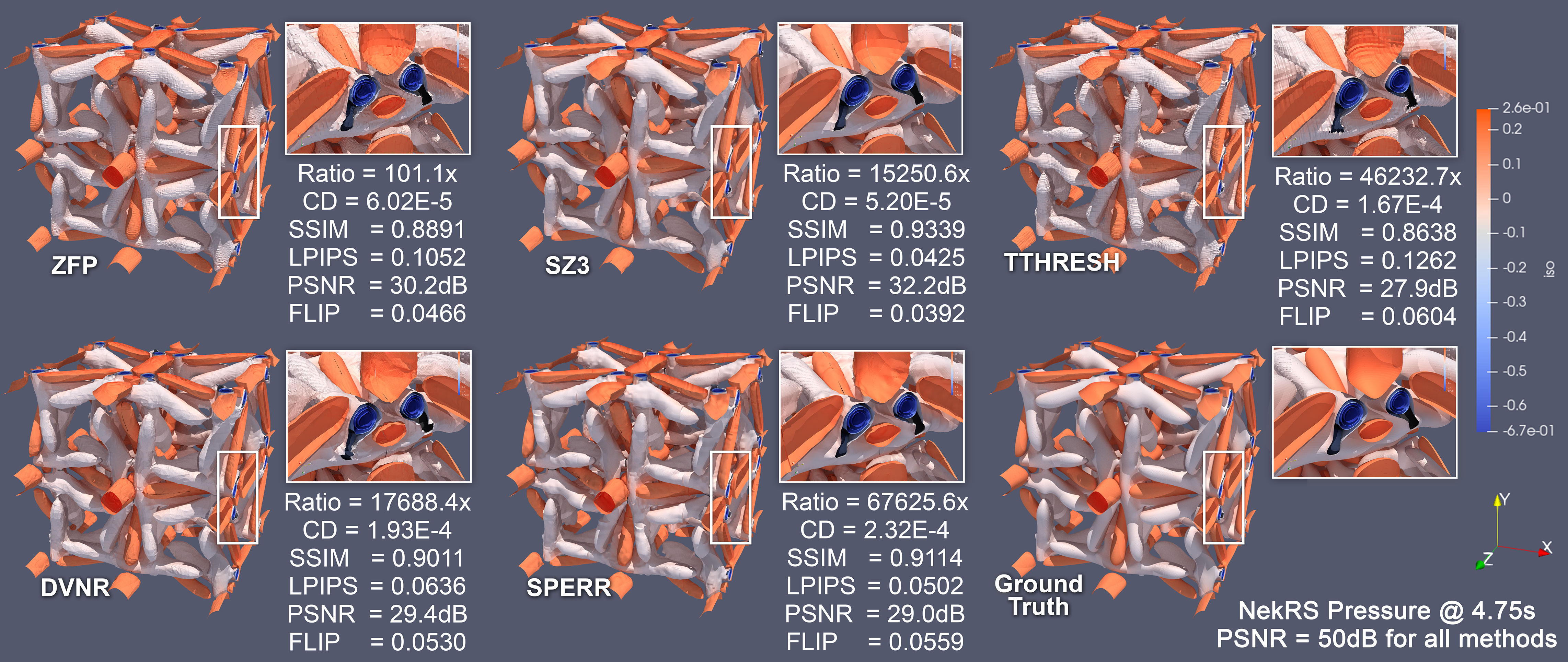}
  \vspace{-2em}
  \caption{\label{fig:compression-artifacts}Isosurfaces extracted from DVNR, the ground truth data and the data compressed by \zfp, \szthree and \tthresh. We targeted all the compressors to PSNR = 50dB. We measured the compression ratio, \tvcg{visualization qualities,} and the bidirectional Chamfer distance (CD)~\cite{barrow1977parametric} between the compressed isosurface and the ground truth. 
  \tvcg{Five isosurfaces are generated using isovalues evenly distributed between the minimum and maximum values of the ground truth data.}
  To highlight compression artifacts, isosurfaces were saved and then rendered with global illumination using ParaView on a single machine.}
  \vspace{-1em}
\end{figure}

Finally, we extracted isosurfaces from DVNR and compared the results with the ground truth and ones derived from other compressors. All compressors were targeted to a PSNR of 50dB, and we measured the compression ratios and isosurface accuracies, gauged by the bidirectional Chamfer Distances (CDs)~\cite{barrow1977parametric} relative to the ground truth \tvcg{isosurfaces}. Our findings are detailed in \Cref{fig:compression-artifacts}.

We found that all compressors introduced artifacts in the extracted isosurfaces. Numerically, the accuracy of isosurfaces produced by DVNR was on par with those produced by \tthresh. \tvcg{Visually, the artifacts in DVNR, \szthree, and \sperr were less prominent and more subtle. \tthresh led in achieving the highest compression ratio for this dataset, followed by DVNR.}

\subsection{Temporal Data Caching Performance}

\begin{figure}[tb]
  \centering
  \vspace{-1em}
  \includegraphics[width=\linewidth]{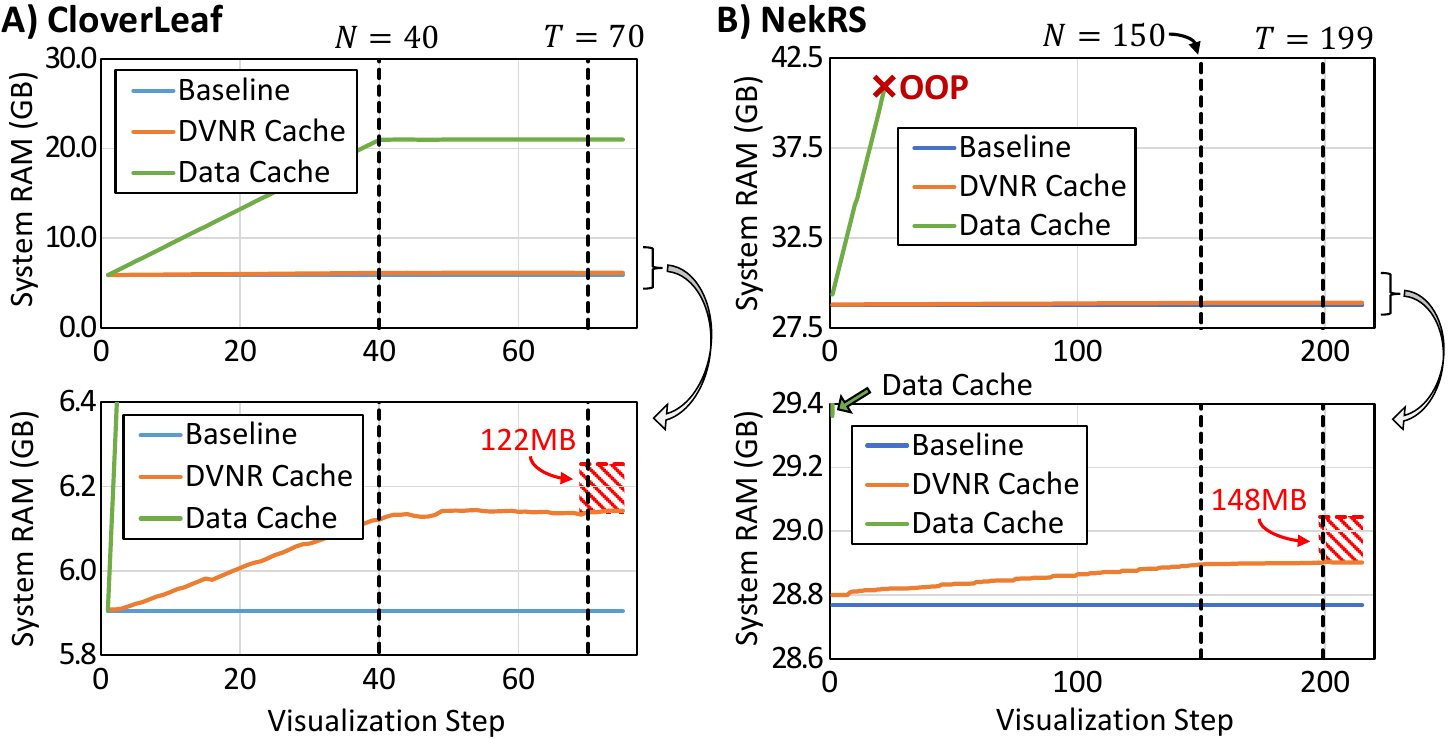}
  \vspace{-2em}
  \caption{\label{fig:case_caching}\tvcg{Memory footprint for temporal data caching using CloverLeaf and NekRS with 4 MPI ranks. For each simulation, direct volume rendering is triggered at the $T^{th}$ visualization step to visualize the most recent $N$ visualization steps. 
  CloverLeaf was configured on a $400^3$ domain, running for 100 visualization steps. 
  NekRS was set on a $420^3$ domain for 220 visualization steps. 
  The red striped lines estimate the additional system memory that would be required to accommodate a fully decoded volume grid if a regular volume rendering method were used. 
  We compared our technique (DVNR Cache) with an method that cached uncompressed data (Data Cache) and a baseline that neither cache data nor render data (Baseline).}}
  \vspace{-0.5em}
\end{figure}

In our study, two synthetic cases were developed using CloverLeaf and NekRS to assess the efficacy of the DVNR-based temporal caching technique. In each case, direct volume rendering was initiated at the $T^{th}$ visualization step through an \insitu trigger. Unlike regular methods that only visualize data available in the current step, our approach visualized data from the most recent $N$ steps, enabled by DVNR-based sliding windows (refer to \Cref{sec:temporal_caching}). We monitored system memory usage \tvcg{and visualization processing time} at each step across different ranks. We compared this technique against a na\"ive method that cached uncompressed data directly in memory and a baseline \tvcg{that did not perform any data caching and volume rendering (other data transformation routines were still executed in each visualization step)}. The results are illustrated in \Cref{fig:case_caching}.

The memory footprint of our DVNR-based implementation aligned with expectations. During the initial  $N$ timesteps, we observed a consistent increase in memory usage, attributable to the generation and caching of new DVNR model. Beyond $N$ steps, this growth plateaued as the window operator reached full capacity, necessitating the eviction of older DVNR models. When compared to the approach of caching uncompressed volume data, our method significantly reduced memory demands. Notably, DVNR caching prevented the NekRS case from crashing due to excessive memory consumption. Additionally, our approach consistently maintained a low memory footprint during the visualization generation process, making it a compelling alternative to conventional compression algorithms.

\subsection{Backward Pathline Tracing with DVNR}

Pathlines \tvcg{are} a well-established visualization technique for examining flow patterns in diverse fields. They can be integrated either forward or backward in time, with backward integration being particularly noteworthy for its ability to aid in identifying regions related to detected events and understanding their causal relationships. However, backward integration presents considerable challenges for \insitu visualization, as it necessitates the storage of numerous timesteps preceding the event. This section details our proof-of-concept efforts to address these challenges through the application of DVNR and temporal data caching in backward pathline visualization.

\begin{figure}[tb]
  \centering
  \includegraphics[width=\linewidth]{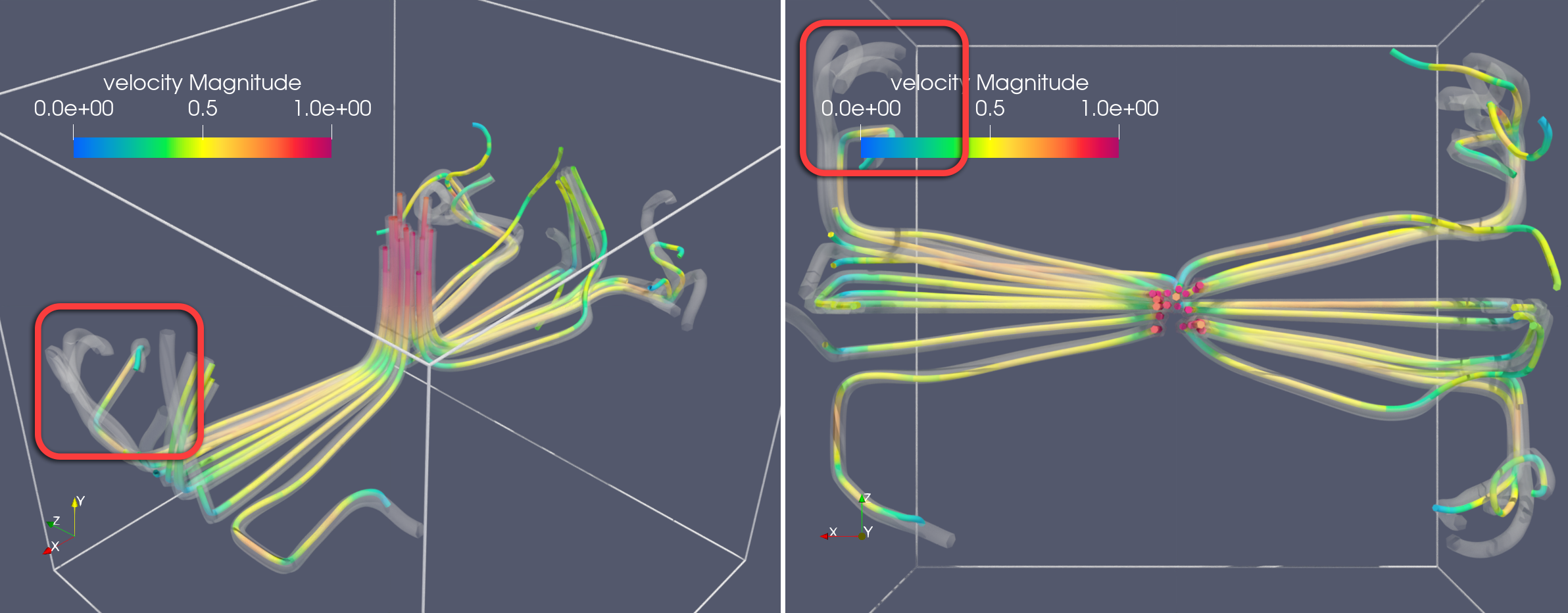}
  \vspace{-2em}
  \caption{\label{fig:rendering-pathline}Comparing the backward pathline tracing results of the ground truth data (transparent tubes) and distributed neural representation (solid tubes). The ground truth tracing was done \posthoc. The pathline tracing implementation directly outputs line data to disks. Renderings were generated using ParaView.}
  \vspace{-1em}
\end{figure}

For simplicity and consistency, we employed the same trigger condition as in the previous section, and constructed a lengthy temporal sliding window using DVNR. Upon trigger activation, we reversed and negated the window, then performed pathline tracing over it, beginning with pre-defined seed points. The resulting pathlines were archived on disk for \posthoc visualization and analysis.

We employed VTKm's pathline tracer and integrated it with DVNR. We calculated backward tracing results for both DVNR and ground truth data. Note that  computing backward pathlines for the ground truth data required storing all the volume data on disk, which was significantly less efficient than our DVNR-based approach. Renderings of generated pathlines are presented in \Cref{fig:rendering-pathline}.

The pathlines generated from DVNRs exhibited noticeable deviations from the ground truth after many integration steps. Further analysis indicated that these deviations were predominantly associated with regions of low velocity magnitudes. This outcome was anticipated, as compression errors in these areas tend to have a more significant impact on integration accuracy. Once these deviations appeared, correcting them proved to be a challenge.

\section{Ablation and Hyperparameter Study}\label{dvnr:sec:ablation}

The behaviors of DVNR's neural network architecture have been extensively studied in existing literature. 
\tvcg{Therefore, in this study, we concentrate on understanding the impacts of three newly introduced optimization techniques. We do this by examining how the overall compression quality changes when each optimization technique is disabled. This type of analysis is known as an ablation study. Additionally, we investigate the choice of hyperparameters for each technique using the same experimental setup.}

\begin{figure}[tb]
    \centering
    \vspace{-0.25em} %
    \includegraphics[width=0.95\linewidth]{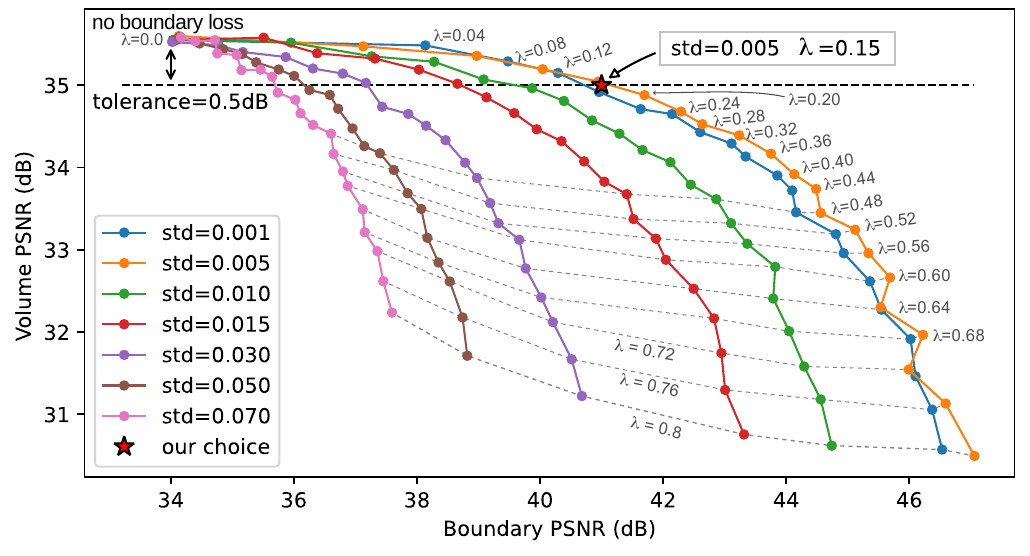}
    \vspace{-1.4em}
    \caption{\label{fig:weight-study}\tvcg{This plot depicts the impact of the weighting factor $\lambda$ on boundary connectivity and overall reconstruction quality. The blue curve represents the average image PSNR of two boundary slices relative to the ground truth slice, while the orange curve illustrates the average volume PSNR of two partitions relative to the ground truth volume. This experiment used the heat release field from the S3D simulation.}}
    \vspace{-0.5em}
\end{figure}

\begin{figure}[tb]
  \centering
  \includegraphics[width=\linewidth]{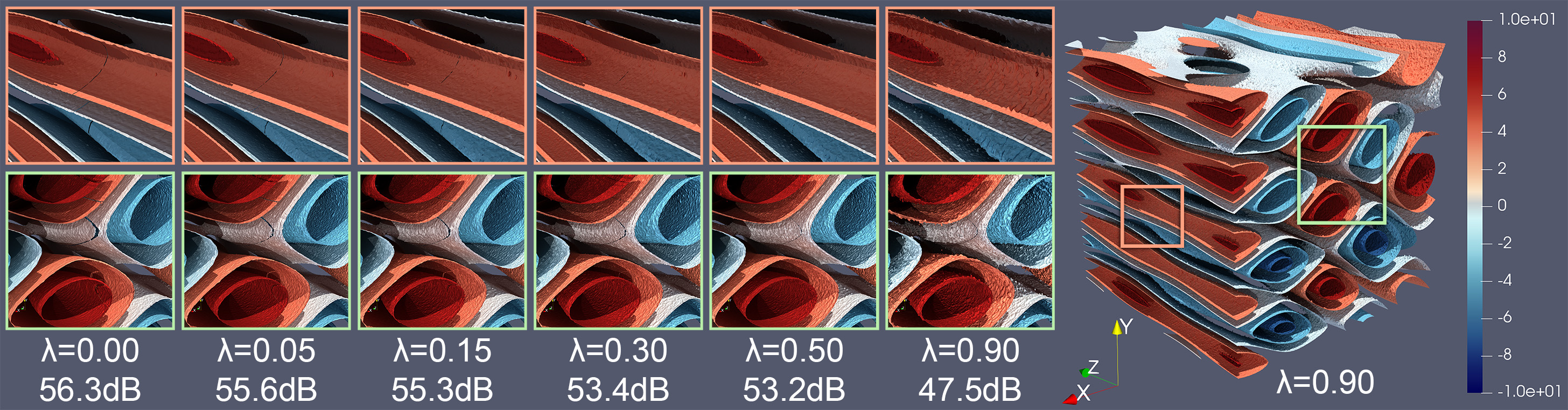}
  \vspace{-2em}
  \caption{\label{fig:contour-weights}Isosurface visualizations with different boundary loss weightings. We used $\sigma=0.005$ in this experiment. \tvcg{Five isosurfaces were generated using isovalues evenly distributed between the minimum and maximum values of the ground truth data.} To highlight artifacts, the isosurfaces were first saved and then rendered with global illumination using ParaView on a single machine.}
  \vspace{-0.5em}
\end{figure}

\subsection{Boundary Loss}\label{dvnr:sec:ablation_boundary_loss}

To study the optimal hyperparameters for boundary loss, we conducted a parameter search using the S3D simulation to assess the effectiveness of the weighting factor $\lambda$ and the standard deviation $\sigma$ in generating boundary samples, as visualized in \Cref{fig:weight-study}. We discovered that incorporating the boundary connectivity loss significantly boosted data accuracy across the partition boundary. However, we also noted that increasing $\lambda$ tended to degrade the overall volume reconstruction quality. Conversely, decreasing $\sigma$ initially led to improvements in boundary accuracy, but continued reduction beyond $\sigma=0.005$ offered no further gains. Based on these insights, we empirically set our volume quality tolerance at 0.5dB and determined that the optimal value for $\lambda$ was $0.15$. \Cref{fig:contour-weights} showcases isosurface visualizations created with various values of $\lambda$.

\subsection{Weight Caching}\label{dvnr:sec:weight_caching}

In the \insitu experiment showcased in \Cref{fig:time_evo}, we also evaluated the performance of DVNR both with and without weight caching. Weight caching offered two primary benefits. Firstly, as previously mentioned, it reduced the DVNR compression time by up to a factor of 10 as the simulation progressed. In the worst-case scenario, weight caching had no negative effects on compression time. Secondly, it enabled the transfer of knowledge from earlier timesteps to current training, enhancing the neural network's reconstruction quality, as demonstrated in \Cref{fig:time_evo}'s PSNR plots. However, weight caching slightly impaired DVNR's compression ratios. This was attributable to two factors. First, initializing DVNR with random weights allowed the optimization process to start at different initial positions, increasing the chance of achieving better compression ratios occasionally. Second, weight caching improved DVNR's accuracy, naturally lowering the overall compression ratio.

\begin{figure*}[tb]
  \centering
  \vspace{-2em}
  \includegraphics[width=0.9\linewidth]{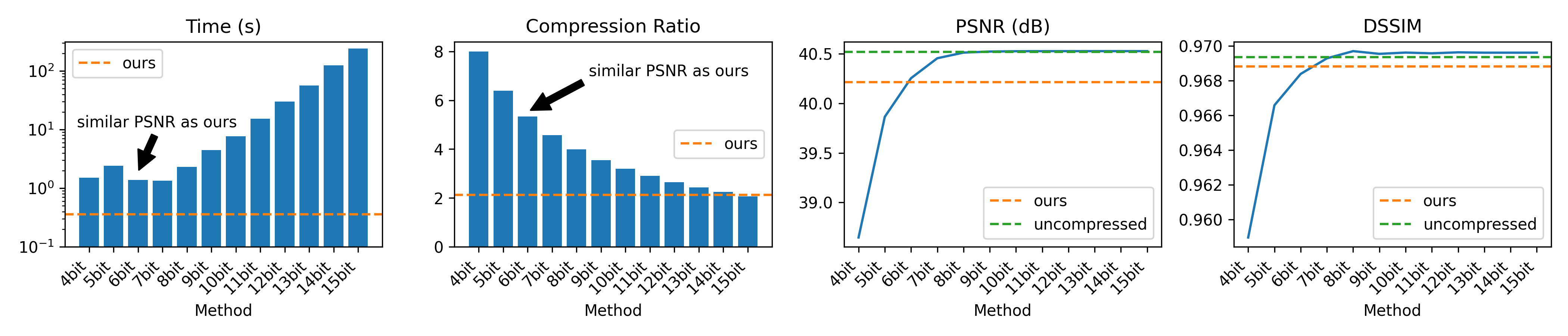}
  \vspace{-1.4em}
  \caption{\label{fig:weight-comp}\tvcg{Comparison of our model compression method with the K-means quantization method implemented by Lu~\etal~\cite{lu2021compressive}. We pretrained a DVNR model on the \textbf{Mechanical Hand} dataset and evaluated the compression ratio, compression time, and model accuracy. Dashed lines indicate the measurements for our method and the uncompressed model.}}
  \vspace{-1.5em}
\end{figure*}

\begin{table}[tb]
\vspace{-0.5em}
\setlength{\tabcolsep}{6pt}
\caption{\tvcg{Comparison of average model compression performance in terms of compression 
ratios (CRs) and PSNR/SSIM changes. Negative $\Delta$ represents a decrease in data reconstruction quality after model compression.
Unaveraged measurements are available in the supplementary material.}}
\label{tab:model_comp_ratios}
\vspace{-1em}
\scriptsize\centering%
\begin{tabu}{cc cccc}
\toprule 
  \multicolumn{2}{c}{} 
& \multicolumn{4}{c}{{
    Model Compression
}} \\
\cmidrule(lr){3-6}
\multicolumn{2}{c}{Dataset} & CR & $\Delta$PSNR & $\Delta$SSIM & $\Delta$DSSIM \\
\midrule
\multicolumn{2}{r}{Magnetic         } & 2.37 & -0.25 & -0.0008 & -0.0054 \\
\multicolumn{2}{r}{Rayleigh Taylor  } & 2.17 & -0.21 & -0.0020 & -0.0047 \\
\multicolumn{2}{r}{Richtmyer Meshkov} & 2.72 & -0.09 & -0.0005 & -0.0030 \\
\multicolumn{2}{r}{S3D H2           } & 2.00 & -0.45 & -0.0024 & -0.0091 \\
\multicolumn{2}{r}{Pawpawsaurus     } & 2.17 & -0.12 & -0.0003 & -0.0018 \\
\multicolumn{2}{r}{Chameleon        } & 3.02 & -0.25 & -0.0003 & -0.0005 \\
\multicolumn{2}{r}{Beechnut         } & 2.40 & -0.01 & -0.0001 & -0.0001 \\
\multicolumn{2}{r}{Tortoise         } & 2.30 & -0.16 & -0.0002 & -0.0006 \\
\midrule  
\multirow{4}{*}{ \begin{tabular}{@{}c@{}}S3D \\ (\insitu)\end{tabular} }
& NH3  & 4.42 & -1.50 & -0.0007  & 0.0492 \\
& O2   & 3.48 & -0.66 & -0.0006  & 0.1368 \\
& N2   & 3.39 & -0.68 & -0.0005  & 0.0325 \\
& Temp & 3.41 & -0.47 & -0.0005  & 0.1437 \\
\midrule
NekRS (\insitu)
& VelMag & 2.15 & -0.29 & -0.0001 & 0.0007 \\
\bottomrule
\end{tabu}
\vspace{-2em}
\end{table}

\subsection{Model Compression}\label{dvnr:sec:weight_compression}

During all our experiments, we also recorded the compression ratios and inaccuracies resulting from model compression. \tvcg{Inaccuracies were measured in terms of PSNR, SSIM and DSSIM changes (\ie $\Delta$PSNR, $\Delta$SSIM, \tvcg{and $\Delta$DSSIM}). In \Cref{fig:time_evo}, we visualize the overall compression ratios before (black dashed lines) and after (blue and orange lines) performing model compressions.}

\tvcg{Qualitatively, our results suggest that even light application of model compression can significantly improve the overall compression ratio, with only a minimal decrease in model accuracy. Interestingly, we observed rare instances where some of the reconstruction metrics improved slightly after model compression. We attribute this to the unintended smoothing of weight values and reduction of fluctuations by lossy compressors, which can occasionally have beneficial effects.}

Quantitatively, we computed the average CRs, $\Delta$PSNRs, $\Delta$SSIMs, \tvcg{and $\Delta$DSSIMs} for the datasets highlighted in both  \Cref{fig:time_evo} and \Cref{fig:static_volume}. For \posthoc evaluations, measurements were calculated by averaging over different model configurations. For \insitu evaluations, corresponding values were averaged over different timesteps. These results are presented in \Cref{tab:model_comp_ratios}.
It is important to note that computing averages over different timesteps or model configurations is not entirely meaningful; thus, we also report unaveraged results in the supplementary material. On average, additional model compression could further reduced the model size by $2-4.5\times$, with losses in PSNR, SSIM, \tvcg{and DSSIM} in less than 2dB, 0.3\% \tvcg{and 15\%}, respectively.

\tvcg{Given the limited time budget for \insitu compression, we prioritized compression speed when selecting the model compression method. This led us to leverage well-optimized floating point compressors. However, exploring the potential of other model compression techniques remains intriguing.

In this spirit, we compared our method with the popular K-means model quantization method~\cite{han2015deep}. Lu~\etal~\cite{lu2021compressive} used this method to compress INRs, but they only applied quantization to the hidden weights of the MLP. We extended their approach to INRs with positional encoding weights by separately applying quantization to each encoding layer. First, we flattened these weights into an 1D array, then clustered them using K-means, and finally quantized them using the cluster centers. We stored the cluster centers and the integer label for each weight. The number of clusters was computed as $2^{B}$, where $B$ is the number of bits we used to store each label.
In our experiment, we varied $B$ and report the model accuracy using quantized weights as well as the compression time and ratio. Our results are shown in  \Cref{fig:weight-comp}. To provide a more insightful comparison, we used the state-of-the-art GPU-accelerated K-means implementation provided by 
CUML.

Our results show that K-means quantization can achieve a better compression ratio and higher model accuracy simultaneously, but at the cost of significantly longer compression time, even with GPU acceleration. We believe K-means quantization can be a promising alternative model compression method for DVNR. However, it still requires further performance optimization to be more competitive against traditional compressors and additional system engineering effort to be easily integrated into an \insitu visualization pipeline. Therefore, we leave it as a promising direction for future work.}

\section{Discussion and Future Work}

We conducted a comprehensive evaluation of DVNR, focusing on its scalability, performance, and usability.

In terms of quality, our DVNR-based technique offers a simple yet effective approach to employing INR for distributed data. We found that DVNR achieved competitive reconstruction quality while being as fast as state-of-the-art traditional compressors. Using ghost region information and introducing a weighted boundary loss allowed us to completely avoid interprocess communication while ensuring reasonable boundary connectivity. With model compression, we further improved the overall compression ratio by $2-4.5$ times, with negligible computational overheads and less than 2dB loss in accuracy on average. Implementing learned network initialization via weight caching also effectively reduced DVNR compression time for \insitu simulations, boosting reconstruction quality. 

In terms of scalability and performance, by leveraging well-established parallel rendering pipelines and the recently proposed INR rendering algorithms, our technique enabled efficient and scalable visualization and analysis. DVNR also exhibited excellent scalability, with weak scaling results approaching ideal efficiencies.

In terms of usability, our technique significantly reduced data size, enabling aggressive temporal data caching. This is essential for achieving the full potential of reactive programming in writing adaptive \insitu workflows. Our case study results demonstrated DVNR's capability to provide the required quality for scalar field visualizations such as volume rendering and isosurface extraction. However, for tasks more sensitive to errors, such as vector field integrations, DVNR still required improvements. Finally, our integration with Ascent was flexible and easy to use, allowing easy achievement of complex tasks like temporal data caching and backward data analysis using \diva's reactive interface.

DVNR exhibits three major limitations. Firstly, INR-based data compression methods, including DVNR, cannot automatically adjust model size to achieve specific accuracy levels, often requiring pilot studies with models of varying sizes. Developing automatic control over INR architecture is a crucial future research direction. Secondly, more DVNR-compatible visualization algorithms need to be developed for highly efficient \insitu visualization and analysis. 
\tvcgm{Thirdly, the base network's architecture still has room for improvement.
DVNR focuses on minimizing the overall training and visualization costs to meet the time budget of \insitu visualization. Consequently, it utilizes the best-performing INR~\cite{muller2022instant} and compatible visualization algorithms~\cite{wu2022instant}. Adapting to more advanced INRs (\eg APMGSRN~\cite{wurster2023adaptively}) with highly optimized implementations remains a future research direction.}

\section{Conclusion}

In this work, we present a novel design for constructing an implicit neural representation of distributed volume data (DVNR) with three specific optimizations for \insitu visualization. We incorporate DVNR into the \diva reactive programming system and enable efficient temporal data caching for \diva. To demonstrate the enhancement, we also integrate it into the Ascent \insitu visualization and analysis infrastructure. Our comprehensive experimental study indicates that DVNR not only offers competitive compression quality but also excels in scalability and usability, making it a feasible solution for many \insitu visualization tasks.

Our design enables memory-efficient reactive programming, which takes an essential step towards realizing the full potential of reactive programming for adaptive \insitu visualization and analysis. We are optimistic that our contributions will inspire further research to advance the state-of-the-art \insitu data processing technology at extreme-scale.

\section*{Acknowledgments}
\noindent This research was supported by the Exascale Computing Project (17-SC-20-SC), a collaborative effort of the U.S. Department of Energy Office of Science and the National Nuclear Security Administration.
This research was also supported in part by the Department of Energy through grant DE-SC0019486.
This research used resources of the Argonne Leadership Computing Facility, which is a DOE Office of Science User Facility supported under Contract DE-AC02-06CH11357.
The authors also express sincere gratitude to Saumil Patel (Argonne National Laboratory), Abhishek Yenpure (Kitware), Jacqueline H. Chen (Sandia National Laboratory), and Martin Rieth (Sandia National Laboratory) for their invaluable assistance and insightful discussions.

\bibliographystyle{IEEEtran}
\bibliography{main}

\begin{thebibliography}{10}
\providecommand{\url}[1]{#1}
\csname url@samestyle\endcsname
\providecommand{\newblock}{\relax}
\providecommand{\bibinfo}[2]{#2}
\providecommand{\BIBentrySTDinterwordspacing}{\spaceskip=0pt\relax}
\providecommand{\BIBentryALTinterwordstretchfactor}{4}
\providecommand{\BIBentryALTinterwordspacing}{\spaceskip=\fontdimen2\font plus
\BIBentryALTinterwordstretchfactor\fontdimen3\font minus
  \fontdimen4\font\relax}
\providecommand{\BIBforeignlanguage}[2]{{%
\expandafter\ifx\csname l@#1\endcsname\relax
\typeout{** WARNING: IEEEtran.bst: No hyphenation pattern has been}%
\typeout{** loaded for the language `#1'. Using the pattern for}%
\typeout{** the default language instead.}%
\else
\language=\csname l@#1\endcsname
\fi
#2}}
\providecommand{\BIBdecl}{\relax}
\BIBdecl

\bibitem{lu2021compressive}
Y.~Lu, K.~Jiang, J.~A. Levine, and M.~Berger, ``Compressive neural
  representations of volumetric scalar fields,'' \emph{Computer Graphics
  Forum}, vol.~40, no.~3, pp. 135--146, 2021.

\bibitem{wu2022instant}
Q.~Wu, D.~Bauer, M.~J. Doyle, and K.-L. Ma, ``Interactive volume visualization
  via multi-resolution hash encoding based neural representation,'' \emph{IEEE
  Transactions on Visualization and Computer Graphics}, vol.~30, no.~8, pp.
  5404--5418, 2024.

\bibitem{9966405}
Q.~Wu, J.~A. Insley, V.~A. Mateevitsi, S.~Rizzi, and K.-L. Ma, ``Distributed
  volumetric neural representation for in situ visualization and analysis,'' in
  \emph{2022 IEEE 12th Symposium on Large Data Analysis and Visualization
  (LDAV)}, 2022, pp. 1--2.

\bibitem{InSituTrigger}
M.~Larsen, A.~Woods, N.~Marsaglia, A.~Biswas, S.~Dutta, C.~Harrison, and
  H.~Childs, ``A flexible system for in situ triggers,'' in \emph{Proceedings
  of the Workshop on In Situ Infrastructures for Enabling Extreme-Scale
  Analysis and Visualization (ISAV)}, 2018, pp. 1--6.

\bibitem{wu2020diva}
Q.~Wu, T.~Neuroth, O.~Igouchkine, K.~Aditya, J.~H. Chen, and K.-L. Ma,
  ``{DIVA}: A declarative and reactive language for in situ visualization,'' in
  \emph{2020 IEEE 10th Symposium on Large Data Analysis and Visualization
  (LDAV)}, 2020, pp. 1--11.

\bibitem{larsen2022ascent}
M.~Larsen, E.~Brugger, H.~Childs, and C.~Harrison, ``{Ascent}: A flyweight in
  situ library for exascale simulations,'' in \emph{In Situ Visualization for
  Computational Science}.\hskip 1em plus 0.5em minus 0.4em\relax Springer,
  2022, pp. 255--279.

\bibitem{Zstandard}
``{Zstandard}, fast real-time compression algorithm,''
  \url{https://facebook.github.io/zstd/}, accessed: 2024-01-01.

\bibitem{lindstrom2014fixed}
P.~Lindstrom, ``Fixed-rate compressed floating-point arrays,'' \emph{IEEE
  Transactions on Visualization and Computer Graphics}, vol.~20, no.~12, pp.
  2674--2683, 2014.

\bibitem{pearlman2004efficient}
W.~A. Pearlman, A.~Islam, N.~Nagaraj, and A.~Said, ``Efficient, low-complexity
  image coding with a set-partitioning embedded block coder,'' \emph{IEEE
  Transactions on Circuits and Systems for Video Technology}, vol.~14, no.~11,
  pp. 1219--1235, 2004.

\bibitem{tang2006three}
X.~Tang and W.~A. Pearlman, ``Three-dimensional wavelet-based compression of
  hyperspectral images,'' in \emph{Hyperspectral data compression}.\hskip 1em
  plus 0.5em minus 0.4em\relax Springer, 2006, pp. 273--308.

\bibitem{li2023lossy}
S.~Li, P.~Lindstrom, and J.~Clyne, ``Lossy scientific data compression with
  {SPERR},'' in \emph{2023 IEEE International Parallel and Distributed
  Processing Symposium (IPDPS)}, 2023, pp. 1007--1017.

\bibitem{suter2013tamresh}
S.~K. Suter, M.~Makhynia, and R.~Pajarola, ``{TAMRESH} - tensor approximation
  multiresolution hierarchy for interactive volume visualization,'' in
  \emph{Proceedings of the 15th Eurographics Conference on Visualization
  (EuroVis)}, 2013, pp. 151--160.

\bibitem{ballester2016lossy}
R.~Ballester-Ripoll and R.~Pajarola, ``Lossy volume compression using tucker
  truncation and thresholding,'' \emph{The Visual Computer}, vol.~32, no.~11,
  pp. 1433--1446, 2016.

\bibitem{ballester2019tthresh}
R.~Ballester-Ripoll, P.~Lindstrom, and R.~Pajarola, ``{TTHRESH}: Tensor
  compression for multidimensional visual data,'' \emph{IEEE Transactions on
  Visualization and Computer Graphics}, vol.~26, no.~9, pp. 2891--2903, 2020.

\bibitem{tao2017significantly}
K.~Zhao, S.~Di, X.~Liang, S.~Li, D.~Tao, Z.~Chen, and F.~Cappello,
  ``Significantly improving lossy compression for hpc datasets with
  second-order prediction and parameter optimization,'' in \emph{Proceedings of
  the 29th International Symposium on High-Performance Parallel and Distributed
  Computing (HPDC)}, 2020, pp. 89--100.

\bibitem{liang2018error}
X.~Liang, S.~Di, D.~Tao, S.~Li, S.~Li, H.~Guo, Z.~Chen, and F.~Cappello,
  ``Error-controlled lossy compression optimized for high compression ratios of
  scientific datasets,'' in \emph{2018 IEEE International Conference on Big
  Data (Big Data)}, 2018, pp. 438--447.

\bibitem{zhao2021optimizing}
K.~Zhao, S.~Di, M.~Dmitriev, T.-L.~D. Tonellot, Z.~Chen, and F.~Cappello,
  ``Optimizing error-bounded lossy compression for scientific data by dynamic
  spline interpolation,'' in \emph{2021 IEEE 37th International Conference on
  Data Engineering (ICDE)}, 2021, pp. 1643--1654.

\bibitem{jain2017compressed}
S.~Jain, W.~Griffin, A.~Godil, J.~W. Bullard, J.~Terrill, and A.~Varshney,
  ``Compressed volume rendering using deep learning,'' in \emph{2017 IEEE 7th
  Symposium on Large Data Analysis and Visualization (LDAV)}, 2017, pp.
  1187--1194.

\bibitem{sitzmann2020implicit}
V.~Sitzmann, J.~N.~P. Martel, A.~W. Bergman, D.~B. Lindell, and G.~Wetzstein,
  ``Implicit neural representations with periodic activation functions,'' in
  \emph{Proceedings of the 34th International Conference on Neural Information
  Processing Systems (NIPS)}, vol.~33, 2020, pp. 7462--7473.

\bibitem{he2016deep}
K.~He, X.~Zhang, S.~Ren, and J.~Sun, ``Deep residual learning for image
  recognition,'' in \emph{2016 IEEE Conference on Computer Vision and Pattern
  Recognition (CVPR)}, 2016, pp. 770--778.

\bibitem{weiss2021fast}
S.~Weiss, P.~Herm{\"u}ller, and R.~Westermann, ``Fast neural representations
  for direct volume rendering,'' \emph{Computer Graphics Forum}, vol.~41,
  no.~6, pp. 196--211, 2022.

\bibitem{wurster2023adaptively}
S.~W. Wurster, T.~Xiong, H.-W. Shen, H.~Guo, and T.~Peterka, ``Adaptively
  placed multi-grid scene representation networks for large-scale data
  visualization,'' \emph{IEEE Transactions on Visualization and Computer
  Graphics}, 2023.

\bibitem{han2022coordnet}
J.~Han and C.~Wang, ``{CoordNet}: Data generation and visualization generation
  for time-varying volumes via a coordinate-based neural network,'' \emph{IEEE
  Transactions on Visualization and Computer Graphics}, 2022.

\bibitem{kim2022neuralvdb}
D.~Kim, M.~Lee, and K.~Museth, ``{NeuralVDB}: High-resolution sparse volume
  representation using hierarchical neural networks,'' \emph{ACM Trans.
  Graph.}, vol.~43, no.~2, pp. 1--21, 2024.

\bibitem{rebain2021derf}
D.~Rebain, W.~Jiang, S.~Yazdani, K.~Li, K.~M. Yi, and A.~Tagliasacchi, ``Derf:
  Decomposed radiance fields,'' in \emph{Proceedings of the IEEE/CVF Conference
  on Computer Vision and Pattern Recognition}, 2021, pp. 14\,153--14\,161.

\bibitem{reiser2021kilonerf}
C.~Reiser, S.~Peng, Y.~Liao, and A.~Geiger, ``{KiloNeRF}: Speeding up neural
  radiance fields with thousands of tiny mlps,'' in \emph{Proceedings of the
  IEEE/CVF international conference on computer vision}, 2021, pp.
  14\,335--14\,345.

\bibitem{saragadam2022miner}
V.~Saragadam, J.~Tan, G.~Balakrishnan, R.~G. Baraniuk, and A.~Veeraraghavan,
  ``Miner: Multiscale implicit neural representation,'' in \emph{European
  Conference on Computer Vision}, 2022, pp. 318--333.

\bibitem{tang2024ecnr}
K.~Tang and C.~Wang, ``Ecnr: efficient compressive neural representation of
  time-varying volumetric datasets,'' in \emph{2024 IEEE 17th Pacific
  Visualization Conference (PacificVis)}, 2024, pp. 72--81.

\bibitem{muller2022instant}
T.~M{\"u}ller, A.~Evans, C.~Schied, and A.~Keller, ``Instant neural graphics
  primitives with a multiresolution hash encoding,'' \emph{ACM Trans. Graph.},
  vol.~41, no.~4, pp. 1--15, 2022.

\bibitem{tcnn}
T.~M\"{u}ller, ``tiny-cuda-nn,'' \url{https://github.com/NVlabs/tiny-cuda-nn},
  4 2021, version 1.7. License: BSD-3-Clause.

\bibitem{han2015deep}
S.~Han, H.~Mao, and W.~J. Dally, ``Deep compression: Compressing deep neural
  networks with pruning, trained quantization and huffman coding,'' \emph{arXiv
  preprint arXiv:1510.00149}, 2015.

\bibitem{han2015learning}
S.~Han, J.~Pool, J.~Tran, and W.~Dally, ``Learning both weights and connections
  for efficient neural network,'' \emph{Advances in neural information
  processing systems}, vol.~28, 2015.

\bibitem{frankle2018lottery}
J.~Frankle and M.~Carbin, ``The lottery ticket hypothesis: Finding sparse,
  trainable neural networks,'' \emph{arXiv preprint arXiv:1803.03635}, 2018.

\bibitem{gale2019state}
T.~Gale, E.~Elsen, and S.~Hooker, ``The state of sparsity in deep neural
  networks,'' \emph{arXiv preprint arXiv:1902.09574}, 2019.

\bibitem{lin2017towards}
X.~Lin, C.~Zhao, and W.~Pan, ``Towards accurate binary convolutional neural
  network,'' \emph{Advances in neural information processing systems}, vol.~30,
  2017.

\bibitem{tung2018clip}
F.~Tung and G.~Mori, ``Clip-q: Deep network compression learning by in-parallel
  pruning-quantization,'' in \emph{Proceedings of the IEEE conference on
  computer vision and pattern recognition}, 2018, pp. 7873--7882.

\bibitem{girish2023shacira}
S.~Girish, A.~Shrivastava, and K.~Gupta, ``Shacira: Scalable hash-grid
  compression for implicit neural representations,'' in \emph{Proceedings of
  the IEEE/CVF International Conference on Computer Vision}, 2023, pp.
  17\,513--17\,524.

\bibitem{mishra2020survey}
R.~Mishra, H.~P. Gupta, and T.~Dutta, ``A survey on deep neural network
  compression: Challenges, overview, and solutions,'' \emph{arXiv preprint
  arXiv:2010.03954}, 2020.

\bibitem{finn2017model}
C.~Finn, P.~Abbeel, and S.~Levine, ``Model-agnostic meta-learning for fast
  adaptation of deep networks,'' in \emph{International conference on machine
  learning}.\hskip 1em plus 0.5em minus 0.4em\relax PMLR, 2017, pp. 1126--1135.

\bibitem{nichol2018first}
A.~Nichol, J.~Achiam, and J.~Schulman, ``On first-order meta-learning
  algorithms,'' \emph{arXiv preprint arXiv:1803.02999}, 2018.

\bibitem{sitzmann2020metasdf}
V.~Sitzmann, E.~Chan, R.~Tucker, N.~Snavely, and G.~Wetzstein, ``Metasdf:
  Meta-learning signed distance functions,'' \emph{Advances in Neural
  Information Processing Systems}, vol.~33, pp. 10\,136--10\,147, 2020.

\bibitem{tancik2021learned}
M.~Tancik, B.~Mildenhall, T.~Wang, D.~Schmidt, P.~P. Srinivasan, J.~T. Barron,
  and R.~Ng, ``Learned initializations for optimizing coordinate-based neural
  representations,'' in \emph{Proceedings of the IEEE/CVF Conference on
  Computer Vision and Pattern Recognition}, 2021, pp. 2846--2855.

\bibitem{bennett2012combining}
J.~C. Bennett, H.~Abbasi, P.-T. Bremer, R.~Grout, A.~Gyulassy, T.~Jin,
  S.~Klasky, H.~Kolla, M.~Parashar, V.~Pascucci, P.~Pebay, D.~Thompson, H.~Yu,
  F.~Zhang, and J.~Chen, ``{Combining In-Situ and in-Transit Processing to
  Enable Extreme-Scale Scientific Analysis},'' in \emph{{Proc. Int. Conf. SC}},
  2012, pp. 1--9.

\bibitem{Ascent}
M.~Larsen, J.~Ahrens, U.~Ayachit, E.~Brugger, H.~Childs, B.~Geveci, and
  C.~Harrison, ``{The ALPINE In Situ Infrastructure: Ascending from the Ashes
  of Strawman},'' in \emph{{Proc. Workshop ISAV}}, 2017, pp. 42--46.

\bibitem{demarle2021situ}
D.~E. DeMarle and A.~C. Bauer, ``In situ visualization with temporal caching,''
  \emph{Computing in Science \& Engineering}, vol.~23, no.~3, pp. 25--33, 2021.

\bibitem{ParaView}
J.~Ahrens, B.~Geveci, and C.~Law, ``{ParaView: An End-User Tool for Large-Data
  Visualization},'' in \emph{{Visualization Handbook}}.\hskip 1em plus 0.5em
  minus 0.4em\relax Butterworth-Heinemann, 2005, pp. 717--731.

\bibitem{291528}
S.~Molnar, M.~Cox, D.~Ellsworth, and H.~Fuchs, ``A sorting classification of
  parallel rendering,'' \emph{IEEE Computer Graphics and Applications},
  vol.~14, no.~4, pp. 23--32, 1994.

\bibitem{diskin2011comparison}
B.~Diskin and J.~L. Thomas, ``Comparison of node-centered and cell-centered
  unstructured finite-volume discretizations: inviscid fluxes,'' \emph{AIAA
  journal}, vol.~49, no.~4, pp. 836--854, 2011.

\bibitem{liang2022sz3}
X.~Liang, K.~Zhao, S.~Di, S.~Li, R.~Underwood, A.~M. Gok, J.~Tian, J.~Deng,
  J.~C. Calhoun, D.~Tao \emph{et~al.}, ``Sz3: A modular framework for composing
  prediction-based error-bounded lossy compressors,'' \emph{IEEE Transactions
  on Big Data}, vol.~9, no.~2, pp. 485--498, 2022.

\bibitem{wu2023hyperinr}
Q.~Wu, D.~Bauer, Y.~Chen, and K.-L. Ma, ``Hyperinr: A fast and predictive
  hypernetwork for implicit neural representations via knowledge
  distillation,'' \emph{arXiv preprint arXiv:2304.04188}, 2023.

\bibitem{lindstrom2006fast}
P.~Lindstrom and M.~Isenburg, ``Fast and efficient compression of
  floating-point data,'' \emph{IEEE Transactions on Visualization and Computer
  Graphics}, vol.~12, no.~5, pp. 1245--1250, 2006.

\bibitem{Fran}
C.~Elliott and P.~Hudak, ``Functional reactive animation,'' in \emph{Proc. 2nd
  ACM SIGPLAN ICFP}, 1997, pp. 263--273.

\bibitem{wald2016ospray}
I.~Wald, G.~P. Johnson, J.~Amstutz, C.~Brownlee, A.~Knoll, J.~Jeffers,
  J.~G{\"u}nther, and P.~Navr{\'a}til, ``Ospray-a cpu ray tracing framework for
  scientific visualization,'' \emph{IEEE Transactions on Visualization and
  Computer Graphics}, vol.~23, no.~1, pp. 931--940, 2016.

\bibitem{VTKm}
K.~Moreland, C.~Sewell, W.~Usher, L.-T. Lo, J.~Meredith, D.~Pugmire, J.~Kress,
  H.~Schroots, K.-L. Ma, H.~Childs, M.~Larsen, C.-M. Chen, R.~Maynard, and
  B.~Geveci, ``{VTK-m: Accelerating the Visualization Toolkit for Massively
  Threaded Architectures},'' \emph{IEEE Comput. Graph. Appl.}, vol.~36, no.~3,
  pp. 48--58, 2016.

\bibitem{mallinson2013cloverleaf}
\BIBentryALTinterwordspacing
A.~Mallinson, D.~A. Beckingsale, W.~Gaudin, J.~Herdman, J.~M. Levesque, and
  S.~A. Jarvis, ``Cloverleaf: Preparing hydrodynamics codes for exascale,''
  \emph{The Cray User Group}, May 6-9 2013. [Online]. Available:
  \url{https://uk-mac.github.io/CloverLeaf3D/}
\BIBentrySTDinterwordspacing

\bibitem{fischer2022nekrs}
P.~Fischer, S.~Kerkemeier, M.~Min, Y.-H. Lan, M.~Phillips, T.~Rathnayake,
  E.~Merzari, A.~Tomboulides, A.~Karakus, N.~Chalmers \emph{et~al.}, ``Nekrs, a
  gpu-accelerated spectral element navier-stokes solver,'' \emph{Parallel
  Computing}, vol. 114, p. 102982, 2022.

\bibitem{hawkes2007scalar}
E.~R. Hawkes, R.~Sankaran, J.~C. Sutherland, and J.~H. Chen, ``{Scalar mixing
  in direct numerical simulations of temporally evolving plane jet flames with
  skeletal CO/H2 kinetics},'' \emph{Proc. Combust. Inst.}, vol.~31, no.~1, pp.
  1633--1640, 2007.

\bibitem{chen2009terascale}
J.~H. Chen, A.~Choudhary, B.~De~Supinski, M.~DeVries, E.~R. Hawkes, S.~Klasky,
  W.-K. Liao, K.-L. Ma, J.~Mellor-Crummey, N.~Podhorszki \emph{et~al.},
  ``Terascale direct numerical simulations of turbulent combustion using s3d,''
  \emph{Computational Science \& Discovery}, vol.~2, no.~1, p. 015001, 2009.

\bibitem{wang2004image}
Z.~Wang, A.~C. Bovik, H.~R. Sheikh, and E.~P. Simoncelli, ``Image quality
  assessment: from error visibility to structural similarity,'' \emph{IEEE
  transactions on image processing}, vol.~13, no.~4, pp. 600--612, 2004.

\bibitem{baker2022dssim}
A.~H. Baker, A.~Pinard, and D.~M. Hammerling, ``On a structural similarity
  index approach for floating-point data,'' \emph{IEEE Transactions on
  Visualization and Computer Graphics}, pp. 1--13, 2023.

\bibitem{zhang2018perceptual}
R.~Zhang, P.~Isola, A.~A. Efros, E.~Shechtman, and O.~Wang, ``The unreasonable
  effectiveness of deep features as a perceptual metric,'' in \emph{CVPR},
  2018.

\bibitem{Andersson2020}
P.~Andersson, J.~Nilsson, T.~Akenine{-}M{\"{o}}ller, M.~Oskarsson,
  K.~{\AA}str{\"{o}}m, and M.~D. Fairchild, ``{{\FLIP:} {A} Difference
  Evaluator for Alternating Images},'' \emph{Proceedings of the ACM on Computer
  Graphics and Interactive Techniques}, vol.~3, no.~2, pp. 15:1--15:23, 2020.

\bibitem{magnetic_reconnection}
F.~Guo, H.~Li, W.~Daughton, and Y.-H. Liu, ``Formation of hard power laws in
  the energetic particle spectra resulting from relativistic magnetic
  reconnection,'' \emph{Phys. Rev. Lett.}, vol. 113, p. 155005, 2014.

\bibitem{miranda}
A.~W. Cook, W.~Cabot, and P.~L. Miller, ``The mixing transition in
  {R}ayleigh-{T}aylor instability,'' \emph{Journal of Fluid Mechanics}, vol.
  511, pp. 333--362, 2004.

\bibitem{richtmyer_meshkov}
R.~H. Cohen, W.~P. Dannevik, A.~M. Dimits, D.~E. Eliason, A.~A. Mirin, Y.~Zhou,
  D.~H. Porter, and P.~R. Woodward, ``Three-dimensional simulation of a
  richtmyer-meshkov instability with a two-scale initial perturbation,''
  \emph{Physics of Fluids}, vol.~14, no.~10, pp. 3692--3709, 2002.

\bibitem{s3d}
M.~Rieth, A.~Gruber, F.~A. Williams, and J.~H. Chen, ``Enhanced burning rates
  in hydrogen-enriched turbulent premixed flames by diffusion of molecular and
  atomic hydrogen,'' \emph{Combustion and Flame}, p. 111740, 2021.

\bibitem{pawpawsaurus}
\BIBentryALTinterwordspacing
A.~Paulina-Carabajal, Y.-N. Lee, and L.~L. Jacobs, ``Pawpawsaurus campbelli,''
  Digital Morphology, 2016, accessed December 27, 2023. [Online]. Available:
  \url{http://digimorph.org/specimens/Pawpawsaurus_campbelli/}
\BIBentrySTDinterwordspacing

\bibitem{chamaeleo}
\BIBentryALTinterwordspacing
J.~Maisano, ``Chamaeleo calyptratus,'' Digital Morphology, 2003, accessed
  December 27, 2023. [Online]. Available:
  \url{http://digimorph.org/specimens/Chamaeleo_calyptratus/whole/}
\BIBentrySTDinterwordspacing

\bibitem{beechnut}
\BIBentryALTinterwordspacing
{Computer-Assisted Paleoanthropology group} and {Visualization and MultiMedia
  Lab, University of Zurich}, ``A microct scan of a dried beechnut,''
  University of Zurich, 2013, accessed December 27, 2023. [Online]. Available:
  \url{https://www.ifi.uzh.ch/en/vmml/research/datasets.html}
\BIBentrySTDinterwordspacing

\bibitem{tortoise}
\BIBentryALTinterwordspacing
J.~Gray, ``Gopherus polyphemus,'' MorphoSource, 2022, {IP} Holder: Florida
  Museum of Natural History, University of Florida. [Online]. Available:
  \url{https://www.morphosource.org/concern/media/000435676}
\BIBentrySTDinterwordspacing

\bibitem{barrow1977parametric}
H.~G. Barrow, J.~M. Tenenbaum, R.~C. Bolles, and H.~C. Wolf, ``Parametric
  correspondence and chamfer matching: Two new techniques for image matching,''
  in \emph{Proceedings: Image Understanding Workshop}, 1977, pp. 21--27.

\end{thebibliography}

\begin{IEEEbiographynophoto}{Qi Wu,}
a student member of the IEEE,
is a PhD candidate in the VIDI Lab at the University of California, Davis. 
He holds a master's degree in computer graphics from the University of Utah in 2018 
and a bachelor's degree in physics from the Hong Kong University of Science and 
Technology in 2016. His research primarily centers on the advancement of 
hardware-accelerated, machine-learning-augmented visualization techniques in 
order to facilitate the exploration of complex, large-scale scientific applications.
\end{IEEEbiographynophoto}

\begin{IEEEbiographynophoto}{Joseph A. Insley}
is the Team Lead for Visualization and Data Analysis at Argonne National Laboratory's Leadership Computing Facility (ALCF), and Associate Professor in the School of Art and Design at Northern Illinois University (NIU). His research includes the development of parallel and scalable methods for large-scale data analysis and visualization, both post hoc and \insitu, on current and next-generation systems. He investigates the use of advanced display and interaction technologies, including immersive virtual and augmented reality environments, for the visualization and exploration of scientific data.
\end{IEEEbiographynophoto}

\begin{IEEEbiographynophoto}{Victor A. Mateevitsi}
is a Computer Scientist at the Argonne Leadership Computing Facility. His research interests include large scale visualizations, augmented and virtual reality technologies, and novel interaction techniques. He has been a TEDx speaker, has been named one of the ``20 in their 20s'' by Crain's Business Magazine, ``Fifty for the Future®'' by the Illinois Technology Foundation and his work has been featured on popular-press magazines such as Forbes, Popular Mechanics, New Scientist and on pop-culture television programs such as the ``Daily Planet'' and ``All-American Makers''. Victor holds a PhD in Computer Science from the Electronic Visualization Laboratory at the University of Illinois at Chicago.
\end{IEEEbiographynophoto}

\begin{IEEEbiographynophoto}{Silvio Rizzi}
is a Computer Scientist at the Argonne Leadership Computing Facility. His research interests are large-scale and \insitu scientific visualization and analysis, display technologies, and immersive visualization environments. He has an MS in Electrical and Computer Engineering, and a PhD in Industrial Engineering and Operations Research, both from the University of Illinois-Chicago.
\end{IEEEbiographynophoto}

\begin{IEEEbiographynophoto}{Michael E. Papka}
is a senior scientist at Argonne National Laboratory, where he is also deputy associate laboratory director for Computing, Environment and Life Sciences (CELS) and division director of the Argonne Leadership Computing Facility (ALCF). Both his laboratory leadership roles and his research interests relate to high-performance computing in support of scientific discovery. In addition to his duties at Argonne, he is a professor of computer science at University of Illinois Chicago (UIC), where he teaches foundational concepts of computer science and advanced topics in data analytics, data science and high-performance computing. He is a member of the Electronic Visualization Laboratory at UIC.
\end{IEEEbiographynophoto}

\begin{IEEEbiographynophoto}{Kwan-Liu Ma,}
an IEEE Fellow, is a distinguished professor of
computer science at the University of California, Davis. He leads
VIDI Labs and the UC Davis Center of Excellence for Visualization.
Professor Ma received his PhD degree in computer science from
the University of Utah in 1993. His research interests include
visualization, computer graphics, HCI, and HPC. He received
the IEEE VGTC Visualization Technical Achievement Award in
2013 and was inducted to the IEEE Visualization Academy in 2019.
Professor Ma presently serves on the IEEE VIS Steering Committee. 
Contact him via email: ma@cs.ucdavis.edu.
\end{IEEEbiographynophoto}

\vfill

\newpage
\appendix
\section*{Network Configurations}
In this section, we provide a detailed and comprehensive list of all DVNR model configurations that were employed in the experiments discussed. Each configuration is carefully documented to ensure clarity and reproducibility.\\

\noindent
\textbf{\textit{1) Scaling Experiments:}}
The network configurations utilized in the scaling experiment, as highlighted in \Cref{fig:scaling}, are listed as follows:

\begin{lstlisting}
"CloverLeaf": {
  "lrate": 0.005, "lrate_decay": -1, "epochs": 14, 
  "n_neurons": 16,  "n_hidden_layers"     : 2, 
  "n_levels" : 5,   "n_features_per_level": 4, 
  "per_level_scale": 2.0,
  "base_resolution": 8, 
  "log2_hashmap_size": [ 8, 16 ]
}
\end{lstlisting}
\begin{lstlisting}
"NekRS": {
  "lrate": 0.005, "lrate_decay": -1, "epochs": 8, 
  "n_neurons": 16,  "n_hidden_layers"     : 3, 
  "n_levels" : 5,   "n_features_per_level": 4,
  "per_level_scale": 2.0
  "base_resolution": (int)cbrt(1<<log2_hashmap_size), 
  "log2_hashmap_size": [ 10, 16 ]
}
\end{lstlisting}
\begin{lstlisting}
"S3D": {
  "lrate": 0.005, "lrate_decay": -1, "epochs": 16, 
  "n_neurons": 16,  "n_hidden_layers"     : 2, 
  "n_levels" : 4,   "n_features_per_level": 4, 
  "per_level_scale": 2.0
  "base_resolution": (int)cbrt(1<<log2_hashmap_size), 
  "log2_hashmap_size": [ 9, 13 ]
}
\end{lstlisting}

\vspace{1em}\noindent
\textbf{\textit{2) \InSitu Compression Experiments:}}
The network configurations utilized in the compression experiment, as highlighted in \Cref{fig:time_evo}, are listed as follows:

\begin{lstlisting}
"NekRS": {
  "lrate": 0.001, "lrate_decay": -1, "epochs": 4, 
  "n_neurons": 16,  "n_hidden_layers"     : 3, 
  "n_levels" : 5,   "n_features_per_level": 4,
  "per_level_scale": 2.0,
  "base_resolution": (int)cbrt(1<<log2_hashmap_size), 
  "log2_hashmap_size": 12, 
  "target_loss": 0.0105, 
  "zfp_mlp": 0.005, "zfp_enc": 0.010
}
\end{lstlisting}
\begin{lstlisting}
"S3D": {
  "lrate": 0.005, "lrate_decay": -1, "epochs": 16, 
  "n_neurons": 16,  "n_hidden_layers"     : 2, 
  "n_levels" : 4,   "n_features_per_level": 4, 
  "per_level_scale": 2.0, 
  "base_resolution": (int)cbrt(1<<log2_hashmap_size), 
  "log2_hashmap_size": 11, 
  "target_loss": 0.005, 
  "zfp_mlp": 0.01, "zfp_enc": 0.02
}
\end{lstlisting}

\vspace{1em}\noindent
\textbf{\textit{3) Post Hoc Compression Experiments:}}
The network configurations utilized in the compression experiment, as highlighted in \Cref{fig:static_volume}, are listed as follows:

\begin{lstlisting}
"magnetic": [{
  "lrate": 0.004, "lrate_decay": 16, "epochs": 24,
  "n_neurons": 16,  "n_hidden_layers"     : 2,
  "n_levels" : 1,   "n_features_per_level": 4,
  "per_level_scale": 1.5,
  "base_resolution": (int)cbrt(1<<log2_hashmap_size),
  "log2_hashmap_size": [8, 9],
  "zfp_enc": 0.044 "zfp_mlp": 0.022
}, {
  "lrate": 0.004, "lrate_decay": 16, "epochs": 24,
  "n_neurons": 16,  "n_hidden_layers"     : 2,
  "n_levels" : 2,   "n_features_per_level": 8,
  "per_level_scale": 1.5,
  "base_resolution": (int)cbrt(1<<log2_hashmap_size),
  "log2_hashmap_size": [8, 12],
  "zfp_enc": 0.04, "zfp_mlp": 0.02
}, {
  "lrate": 0.004, "lrate_decay": 16, "epochs": 24,
  "n_neurons": 16,  "n_hidden_layers"     : 2,
  "n_levels" : 2,   "n_features_per_level": 8,
  "per_level_scale": 1.5,
  "base_resolution": (int)cbrt(1<<log2_hashmap_size),
  "log2_hashmap_size": [13, 16],
  "zfp_enc": 0.036, "zfp_mlp": 0.018
}]
\end{lstlisting}
\begin{lstlisting}
"rayleigh_taylor": [{
  "lrate": 0.001, "lrate_decay": 16, "epochs": 22,
  "n_neurons": 16,  "n_hidden_layers"     : 4,
  "n_levels" : 1,   "n_features_per_level": 8,
  "per_level_scale": 2.0,
  "base_resolution": (int)cbrt(1<<log2_hashmap_size),
  "log2_hashmap_size": [ 9, 12 ],
  "zfp_enc": 0.005, "zfp_mlp": 0.0025
}, {
  "lrate": 0.001, "lrate_decay": 16, "epochs": 22,
  "n_neurons": 16,  "n_hidden_layers"     : 4,
  "n_levels" : 1,   "n_features_per_level": 8,
  "per_level_scale": 2.0,
  "base_resolution": (int)cbrt(1<<log2_hashmap_size),
  "log2_hashmap_size": [ 13, 16 ],
  "zfp_enc": 0.004, "zfp_mlp": 0.002
}, {
  "lrate": 0.005, "lrate_decay": 22, "epochs": 22,
  "n_neurons": 16,  "n_hidden_layers"     : 4,
  "n_levels" : 1,   "n_features_per_level": 8,
  "per_level_scale": 2.0,
  "base_resolution": (int)cbrt(1<<log2_hashmap_size),
  "log2_hashmap_size": [ 17, 20 ],
  "zfp_enc": 0.003, "zfp_mlp": 0.002
}]
\end{lstlisting}
\begin{lstlisting}
"richtmyer_meshkov": [{
  "lrate": 0.005, "lrate_decay": 22, "epochs": 22,
  "n_neurons": 16,  "n_hidden_layers"     : 2,
  "n_levels" : 1,   "n_features_per_level": 8,
  "per_level_scale": 2.0,
  "base_resolution": (int)cbrt(1<<log2_hashmap_size),
  "log2_hashmap_size": [ 9, 12 ],
  "zfp_enc": 0.030, "zfp_mlp": 0.015
}, {
  "lrate": 0.005, "lrate_decay": 22, "epochs": 22,
  "n_neurons": 16,  "n_hidden_layers"     : 2,
  "n_levels" : 1,   "n_features_per_level": 8,
  "per_level_scale": 2.0,
  "base_resolution": (int)cbrt(1<<log2_hashmap_size),
  "log2_hashmap_size": [ 13, 15 ],
  "zfp_enc": 0.026, "zfp_mlp": 0.013
}, {
  "lrate": 0.005, "lrate_decay": 22, "epochs": 22,
  "n_neurons": 16,  "n_hidden_layers"     : 2,
  "n_levels" : 2,   "n_features_per_level": 8,
  "per_level_scale": 2.0,
  "base_resolution": (int)cbrt(1<<log2_hashmap_size),
  "log2_hashmap_size": [ 15, 18 ],
  "zfp_enc": 0.022, "zfp_mlp": 0.011
}, {
  "lrate": 0.005, "lrate_decay": 22, "epochs": 22,
  "n_neurons": 16,  "n_hidden_layers"     : 2,
  "n_levels" : 2,   "n_features_per_level": 8,
  "per_level_scale": 2.0,
  "base_resolution": (int)cbrt(1<<log2_hashmap_size),
  "log2_hashmap_size": [ 19, 22 ],
  "zfp_enc": 0.02, "zfp_mlp": 0.01
}]
\end{lstlisting}
\begin{lstlisting}
"S3D_H2": [{
  "lrate": 0.005, "lrate_decay": 20, "epochs": 20,
  "n_neurons": 16,  "n_hidden_layers"     : 2,
  "n_levels" : 1,   "n_features_per_level": 4,
  "per_level_scale": 2.0,
  "log2_hashmap_size": [ 6, 9 ],
  "base_resolution": (int)cbrt(1<<log2_hashmap_size),
  "zfp_enc": 0.024, "zfp_mlp": 0.012
}, {
  "lrate": 0.001, "lrate_decay": 24, "epochs": 24,
  "n_neurons": 16,  "n_hidden_layers"     : 2,
  "n_levels" : 1,   "n_features_per_level": 8,
  "per_level_scale": 2.0,
  "log2_hashmap_size": [ 9, 11 ],
  "base_resolution": (int)cbrt(1<<log2_hashmap_size),
  "zfp_enc": 0.010, "zfp_mlp": 0.008
}, {
  "lrate": 0.001, "lrate_decay": 28, "epochs": 28,
  "n_neurons": 16,  "n_hidden_layers"     : 2,
  "n_levels" : 1,   "n_features_per_level": 8,
  "per_level_scale": 2.0,
  "log2_hashmap_size": [ 12, 13 ],
  "base_resolution": (int)cbrt(1<<log2_hashmap_size),
  "zfp_enc": 0.010, "zfp_mlp": 0.008
}, {
  "lrate": 0.001, "lrate_decay": 28, "epochs": 28,
  "n_neurons": 16,  "n_hidden_layers"     : 2,
  "n_levels" : 1,   "n_features_per_level": 8,
  "per_level_scale": 2.0,
  "base_resolution": (int)cbrt(1<<log2_hashmap_size),
  "log2_hashmap_size": [ 15, 16 ],
  "zfp_enc": 0.006, "zfp_mlp": 0.004
}]
\end{lstlisting}
\begin{lstlisting}
"pawpawsaurus": [{
  "lrate": 0.01, "lrate_decay": 32, "epochs": 32,
  "n_neurons": 16,  "n_hidden_layers"     : 1,
  "n_levels" : 1,   "n_features_per_level": 8,
  "per_level_scale": 2.0, 
  "base_resolution": (int)cbrt(1<<log2_hashmap_size),
  "log2_hashmap_size": [ 8, 12 ],
  "zfp_enc": 0.016, "zfp_mlp": 0.008
}, {
  "lrate": 0.01, "lrate_decay": 30, "epochs": 30,
  "n_neurons": 16,  "n_hidden_layers"     : 1,
  "n_levels" : 1,   "n_features_per_level": 8,
  "per_level_scale": 2.0,
  "base_resolution": (int)cbrt(1<<log2_hashmap_size),
  "log2_hashmap_size": [ 13, 17 ],
  "zfp_enc": 0.012, "zfp_mlp": 0.006
}, {
  "lrate": 0.01, "lrate_decay": 20, "epochs": 28,
  "n_neurons": 16,  "n_hidden_layers"     : 2,
  "n_levels" : 2,   "n_features_per_level": 4,
  "per_level_scale": 1.1,
  "base_resolution": 53,
  "log2_hashmap_size": 18,
  "zfp_enc": 0.010, "zfp_mlp": 0.006
}, {
  "lrate": 0.01, "lrate_decay": 20, "epochs": 28,
  "n_neurons": 16,  "n_hidden_layers"     : 2,
  "n_levels" : 3,   "n_features_per_level": 4,
  "per_level_scale": 1.1,
  "base_resolution": 67,
  "log2_hashmap_size": 19,
  "zfp_enc": 0.008, "zfp_mlp": 0.004
}, {
  "lrate": 0.01, "lrate_decay": 20, "epochs": 28,
  "n_neurons": 16,  "n_hidden_layers"     : 2,
  "n_levels" : 3,   "n_features_per_level": 4,
  "per_level_scale": 1.1,
  "base_resolution": 84,
  "log2_hashmap_size": 20,
  "zfp_enc": 0.008, "zfp_mlp": 0.004
}]
\end{lstlisting}
\begin{lstlisting}
"chameleon": [{
  "lrate": 0.008, "lrate_decay": 24, "epochs": 24,
  "n_neurons": 16,  "n_hidden_layers"     : 1,
  "n_levels" : 1,   "n_features_per_level": 4,
  "per_level_scale": 2.0,
  "base_resolution": (int)cbrt(1<<log2_hashmap_size),
  "log2_hashmap_size": [ 7, 14 ],
  "zfp_enc": 0.024, "zfp_mlp": 0.012
}, {
  "lrate": 0.008, "lrate_decay": 24, "epochs": 24,
  "n_neurons": 16,  "n_hidden_layers"     : 2,
  "n_levels" : 2,   "n_features_per_level": 4,
  "per_level_scale": 2.0,
  "base_resolution": (int)cbrt(1<<log2_hashmap_size),
  "log2_hashmap_size": [ 13, 15 ],
  "zfp_enc": 0.020, "zfp_mlp": 0.010
}, {
  "lrate": 0.008, "lrate_decay": 24, "epochs": 24,
  "n_neurons": 16,  "n_hidden_layers"     : 2,
  "n_levels" : 2,   "n_features_per_level": 4,
  "per_level_scale": 2.0,
  "base_resolution": (int)cbrt(1<<log2_hashmap_size),
  "log2_hashmap_size": [ 16, 19 ],
  "zfp_enc": 0.016, "zfp_mlp": 0.008
}]
\end{lstlisting}
\begin{lstlisting}
"beechnut": [{
  "lrate": 0.01, "lrate_decay": 18, "epochs": 24,
  "n_neurons": 16,  "n_hidden_layers"     : 1,
  "n_levels" : 1,   "n_features_per_level": 8,
  "per_level_scale": 1.1,
  "log2_hashmap_size": [ 9, 10 ],
  "base_resolution": (int)cbrt(1<<log2_hashmap_size),
  "zfp_enc": 0.02, "zfp_mlp": 0.01
}, {
  "lrate": 0.01, "lrate_decay": 18, "epochs": 24,
  "n_neurons": 16,  "n_hidden_layers"     : 1,
  "n_levels" : 2,   "n_features_per_level": 8,
  "per_level_scale": 1.1,
  "base_resolution": (int)cbrt(1<<log2_hashmap_size),
  "log2_hashmap_size": [ 10, 17 ],
  "zfp_enc": 0.02, "zfp_mlp": 0.01
}, {
  "lrate": 0.001, "lrate_decay": 16, "epochs": 20,
  "n_neurons": 32,  "n_hidden_layers"     : 4,
  "n_levels" : 4,   "n_features_per_level": 8,
  "per_level_scale": 2.0,
  "base_resolution": 16,
  "log2_hashmap_size": [ 18, 19 ],
  "zfp_enc": 0.004, "zfp_mlp": 0.002
}]
\end{lstlisting}
\begin{lstlisting}
"tortoise": [{
  "lrate": 0.01, "lrate_decay": 28, "epochs": 28,
  "n_neurons": 16,  "n_hidden_layers"     : 2,
  "n_levels" : 1,   "n_features_per_level": 8,
  "per_level_scale": 2.0,
  "base_resolution": (int)cbrt(1<<log2_hashmap_size),
  "log2_hashmap_size": [ 9, 10 ],
  "zfp_enc": 0.026, "zfp_mlp": 0.014
}, {
  "lrate": 0.01, "lrate_decay": 28, "epochs": 28,
  "n_neurons": 16,  "n_hidden_layers"     : 2,
  "n_levels" : 1,   "n_features_per_level": 8,
  "per_level_scale": 2.0,
  "base_resolution": (int)cbrt(1<<log2_hashmap_size),
  "log2_hashmap_size": [ 11, 12 ],
  "zfp_enc": 0.024, "zfp_mlp": 0.012
}, {
  "lrate": 0.01, "lrate_decay": 28, "epochs": 28,
  "n_neurons": 16,  "n_hidden_layers"     : 2,
  "n_levels" : 1,   "n_features_per_level": 8,
  "per_level_scale": 2.0,
  "base_resolution": (int)cbrt(1<<log2_hashmap_size),
  "log2_hashmap_size": [ 13, 14 ],
  "zfp_enc": 0.020, "zfp_mlp": 0.010
}, {
  "lrate": 0.01, "lrate_decay": 28, "epochs": 28,
  "n_neurons": 16,  "n_hidden_layers"     : 2,
  "n_levels" : 1,   "n_features_per_level": 8,
  "per_level_scale": 2.0,
  "base_resolution": (int)cbrt(1<<log2_hashmap_size),
  "log2_hashmap_size": [ 15, 16 ],
  "zfp_enc": 0.018, "zfp_mlp": 0.009
}, {
  "lrate": 0.01, "lrate_decay": 28, "epochs": 28,
  "n_neurons": 16, "n_hidden_layers"     : 3,
  "n_levels" : 2,  "n_features_per_level": 8,
  "per_level_scale": 2.0,
  "base_resolution": (int)cbrt(1<<log2_hashmap_size),
  "log2_hashmap_size": [ 16, 19 ],
  "zfp_enc": 0.018, "zfp_mlp": 0.009
}, {
  "lrate": 0.01, "lrate_decay": 28, "epochs": 28,
  "n_neurons": 16,  "n_hidden_layers"     : 3,
  "n_levels" : 2,   "n_features_per_level": 8,
  "per_level_scale": 2.0,
  "base_resolution": (int)cbrt(1<<log2_hashmap_size),
  "log2_hashmap_size": 20,
  "zfp_enc": 0.014, "zfp_mlp": 0.007
}, {
  "lrate": 0.01, "lrate_decay": 28, "epochs": 28,
  "n_neurons": 16,  "n_hidden_layers"     : 3,
  "n_levels" : 2,   "n_features_per_level": 8,
  "per_level_scale": 2.0, 
  "base_resolution": (int)cbrt(1<<log2_hashmap_size),
  "log2_hashmap_size": [ 21, 22 ],
  "zfp_enc": 0.010, "zfp_mlp": 0.005
},
}
\end{lstlisting}

\vspace{1em}\noindent
\textbf{\textit{4) Volume Rendering Experiments:}}
The network configuration employed in the volume rendering experiment, as illustrated in \Cref{fig:static_volume}, is listed as follows:
\begin{lstlisting}
"params": {
  "n_neurons": 32,  "n_hidden_layers"     : 4
  "n_levels" : 4,   "n_features_per_level": 8,
  "log2_hashmap_size": 19, "base_resolution": 8,
  "per_level_scale": 2.0
}
\end{lstlisting}

\vspace{1em}\noindent
\textbf{\textit{5) Isosurface Extraction Experiments:}}
The network configuration employed in the isosurface extraction experiment, as illustrated in \Cref{fig:compression-artifacts}, is listed as follows:
\begin{lstlisting}
"NekRS": {
  "lrate": 0.01, "lrate_decay": 20, "epochs": 4, 
  "n_neurons": 16,  "n_hidden_layers"     : 1, 
  "n_levels" : 4,   "n_features_per_level": 4, 
  "per_level_scale": 2.0,
  "base_resolution": (int)cbrt(1<<log2_hashmap_size), 
  "log2_hashmap_size": 12, 
  "zfp_mlp": 0.005, "zfp_enc": 0.010
}
\end{lstlisting}

\vspace{1em}\noindent
\textbf{\textit{6) Temporal Caching Experiments:}}
The network configurations employed in the temporal data caching experiment, as illustrated in \Cref{fig:case_caching}, are listed as follows:

\begin{lstlisting}
"CloverLeaf": {
  "epochs": 14, "lrate": 0.01, "lrate_decay": 6,
  "n_neurons": 16,  "n_hidden_layers"     : 1, 
  "n_levels" : 4,   "n_features_per_level": 4, 
  "per_level_scale": 2.0, 
  "base_resolution": (int)cbrt(1<<log2_hashmap_size),
  "log2_hashmap_size": 16
  "zfp_mlp": 0.01, "zfp_enc": 0.02
}
\end{lstlisting}

\begin{lstlisting}
"NekRS": {
  "lrate": 0.01, "lrate_decay": 20, "epochs": 4,
  "n_neurons": 16,  "n_hidden_layers"     : 1, 
  "n_levels" : 4,   "n_features_per_level": 4, 
  "per_level_scale": 2.00,
  "base_resolution": (int)cbrt(1<<log2_hashmap_size), 
  "log2_hashmap_size": 12, 
  "zfp_mlp": 0.005, "zfp_enc": 0.010
}
\end{lstlisting}

\vspace{1em}\noindent
\textbf{\textit{7) Ablation Study:}}
The network configuration employed in the ablation study, as illustrated in \Cref{fig:weight-study}, is listed as follows:

\begin{lstlisting}
"params": {
  "n_neurons": 64,  "n_hidden_layers": 3,  
  "n_levels" : 10,  "n_features_per_level": 8, 
  "log2_hashmap_size": 19, "base_resolution": 4, 
  "per_level_scale": 2.0
}
\end{lstlisting}

\begin{figure*}[b]
  \centering
  \includegraphics[width=\linewidth]{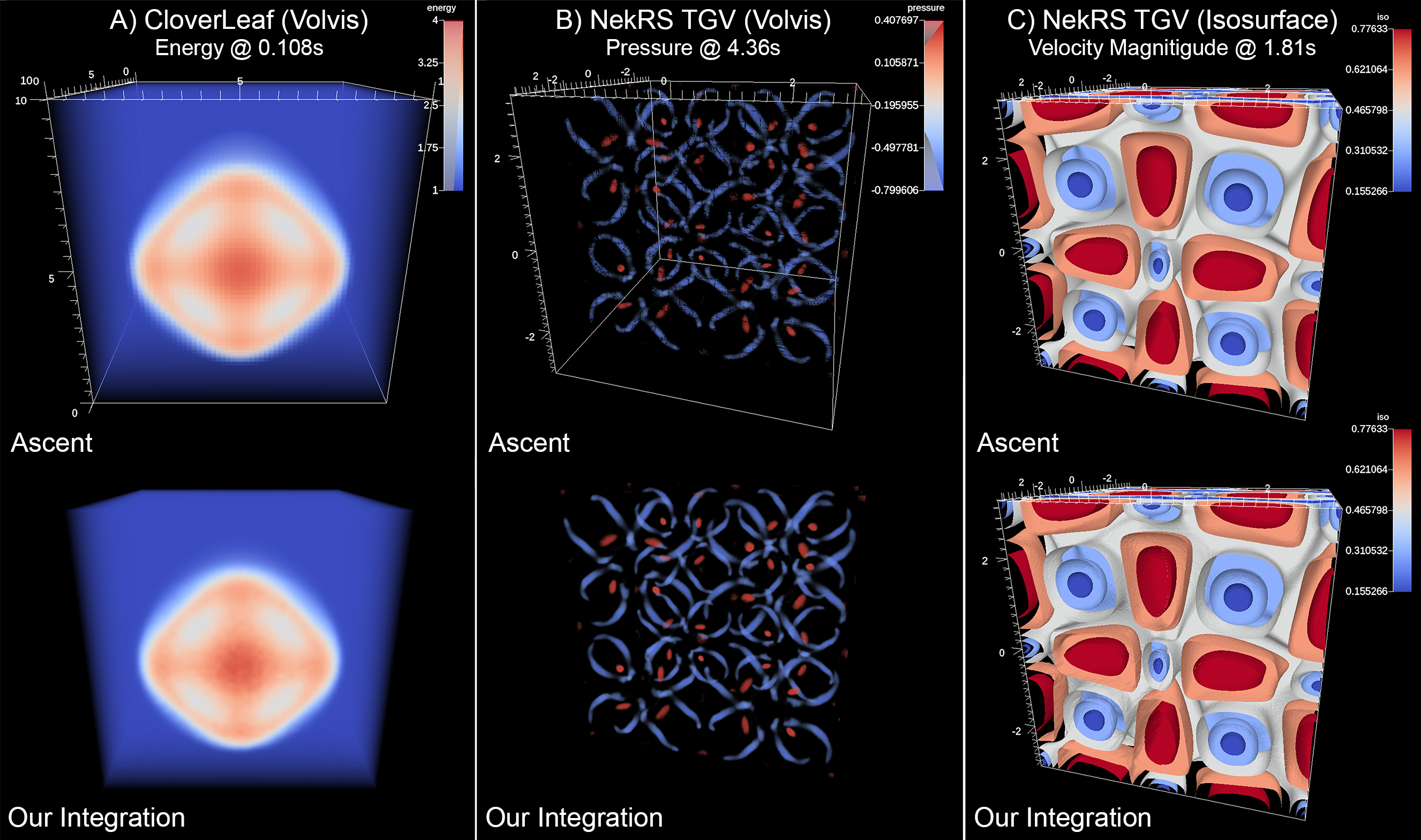}
  \vspace{-2em}
  \caption{\label{fig:rendering-demo}Visualization results generated by our \diva-Ascent integration. \tvcg{Top row: Visualizations of uncompressed data rendered directly by unmodified Ascent. Bottom row: Visualizations of DVNR-compressed data generated by our DIVA-Ascent integration. Columns: A and B) volume renderings of CloverLeaf and NekRS data, respectively; C) isosurface visualizations of NekRS data.}}
  \vspace{-1em}
\end{figure*}

\section*{Extra Results}
In this section, we present additional experimental results and high-resolution images that could not be included in the main text due to space constraints.\\

\begin{table}[h]
\setlength{\tabcolsep}{5pt}
\caption{Comparison of normalized mean squared root error (NRMSE) across 
datasets. Measured values are averaged over timesteps for runs in \Cref{fig:time_evo}, 
and averaged over network configurations for datasets in \Cref{fig:static_volume}.}
\vspace{-1em}
\scriptsize\centering%
\begin{tabu}{cc ccccc}
\toprule 
  \multicolumn{2}{c}{} 
& \multicolumn{5}{c}{{
    Normalized Mean Squared Root Error (NRMSE)
}} \\
\cmidrule(lr){3-7}
\multicolumn{2}{c}{Dataset} &
\multicolumn{1}{r}{DVNR} & 
\multicolumn{1}{c}{\zfp} & 
\multicolumn{1}{r}{\szthree} & 
\multicolumn{1}{r}{\tthresh} &
\multicolumn{1}{r}{\sperr} \\
\midrule
\multicolumn{2}{r}{Magnetic~\cite{magnetic_reconnection}      } & 0.0069 & 0.0130 & 0.0065 & 0.0195 & 0.0044  \\
\multicolumn{2}{r}{Rayleigh Taylor~\cite{miranda}             } & 0.0248 & 0.0240 & 0.0263 & 0.0099 & 0.0045  \\
\multicolumn{2}{r}{Richtmyer Meshkov~\cite{richtmyer_meshkov} } & 0.0357 & 0.0151 & 0.0175 & 0.0092 & 0.0084  \\
\multicolumn{2}{r}{S3D H2~\cite{s3d}                          } & 0.0435 & 0.0216 & 0.0353 & 0.0435 & 0.0054  \\
\multicolumn{2}{r}{Pawpawsaurus~\cite{pawpawsaurus}           } & 0.0138 & 0.0117 & 0.0119 & 0.0138 & 0.0053  \\
\multicolumn{2}{r}{Chameleon~\cite{chamaeleo}                 } & 0.0120 & 0.0048 & 0.0041 & 0.0111 & 0.0035  \\
\multicolumn{2}{r}{Beechnut~\cite{beechnut}                   } & 0.0312 & 0.0318 & 0.0225 & 0.0471 & 0.0164  \\
\multicolumn{2}{r}{Tortoise~\cite{tortoise}                   } & 0.0110 & 0.0142 & 0.0157 & 0.0331 & 0.0048  \\
\midrule  
\multirow{4}{*}{ \begin{tabular}{@{}c@{}}S3D \\ (\insitu)\end{tabular} }
& NH3  & 0.0100 & 0.0071 & 0.0043 & 0.3308 & 0.0020  \\
& O2   & 0.0080 & 0.0066 & 0.0033 & 0.0786 & 0.0015  \\
& N2   & 0.0079 & 0.0067 & 0.0030 & 0.4523 & 0.0013  \\
& Temp & 0.0077 & 0.0067 & 0.0033 & 0.0086 & 0.0016  \\
\midrule
NekRS (\insitu)
& VelMag & 0.0153 & 0.0122 & 0.0068 & 0.0111 & 0.0021 \\
\bottomrule
\end{tabu}
\end{table}

\onecolumn
\begin{figure}[h]
  \centering
  \includegraphics[width=\linewidth]{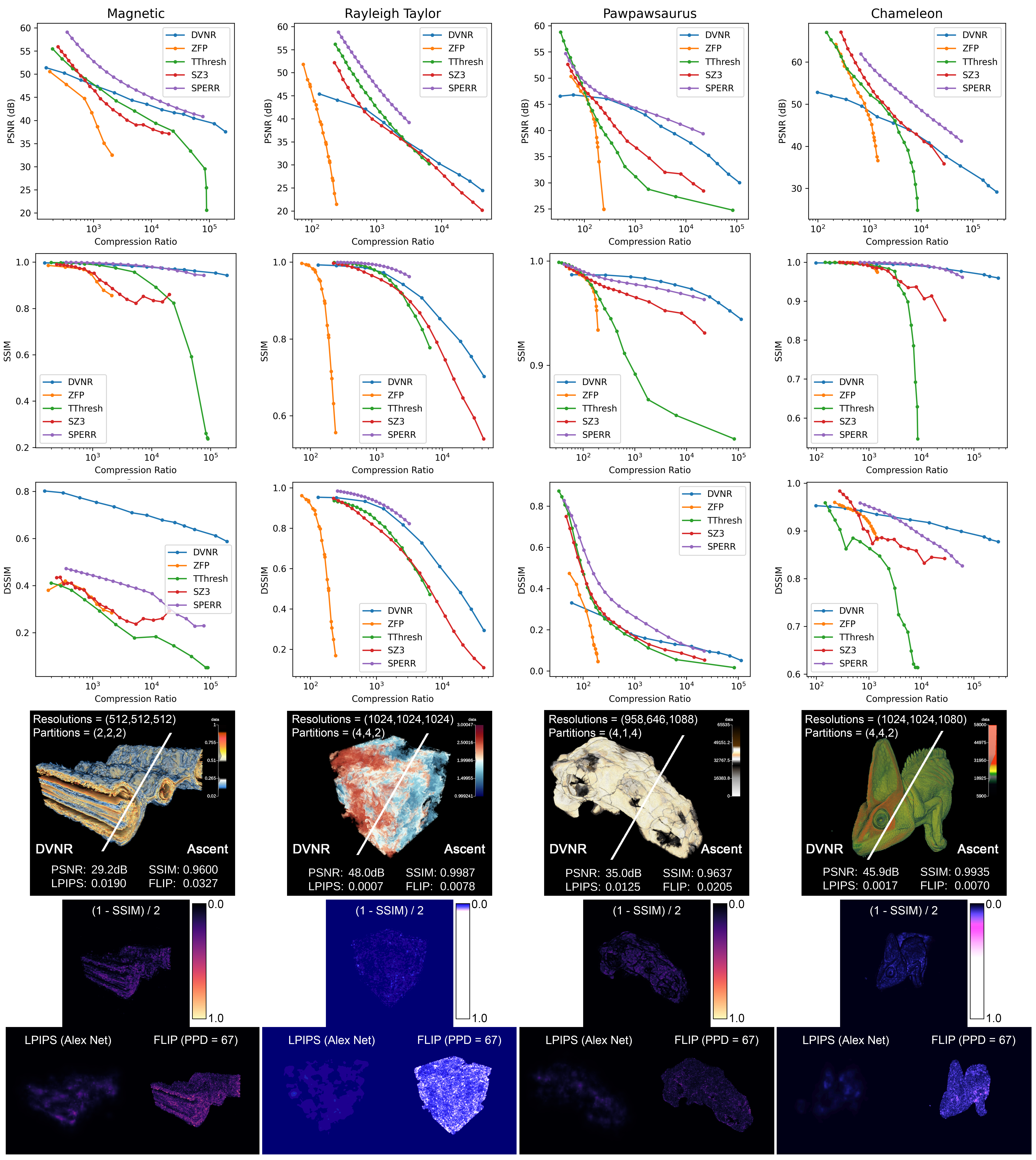}
  \vspace{-2em}
  \caption{Extra results for experiments reported in \Cref{fig:static_volume}.}
  \label{fig:static_volume_more0}
\end{figure}

\onecolumn
\begin{figure}[h]
  \centering
  \includegraphics[width=\linewidth]{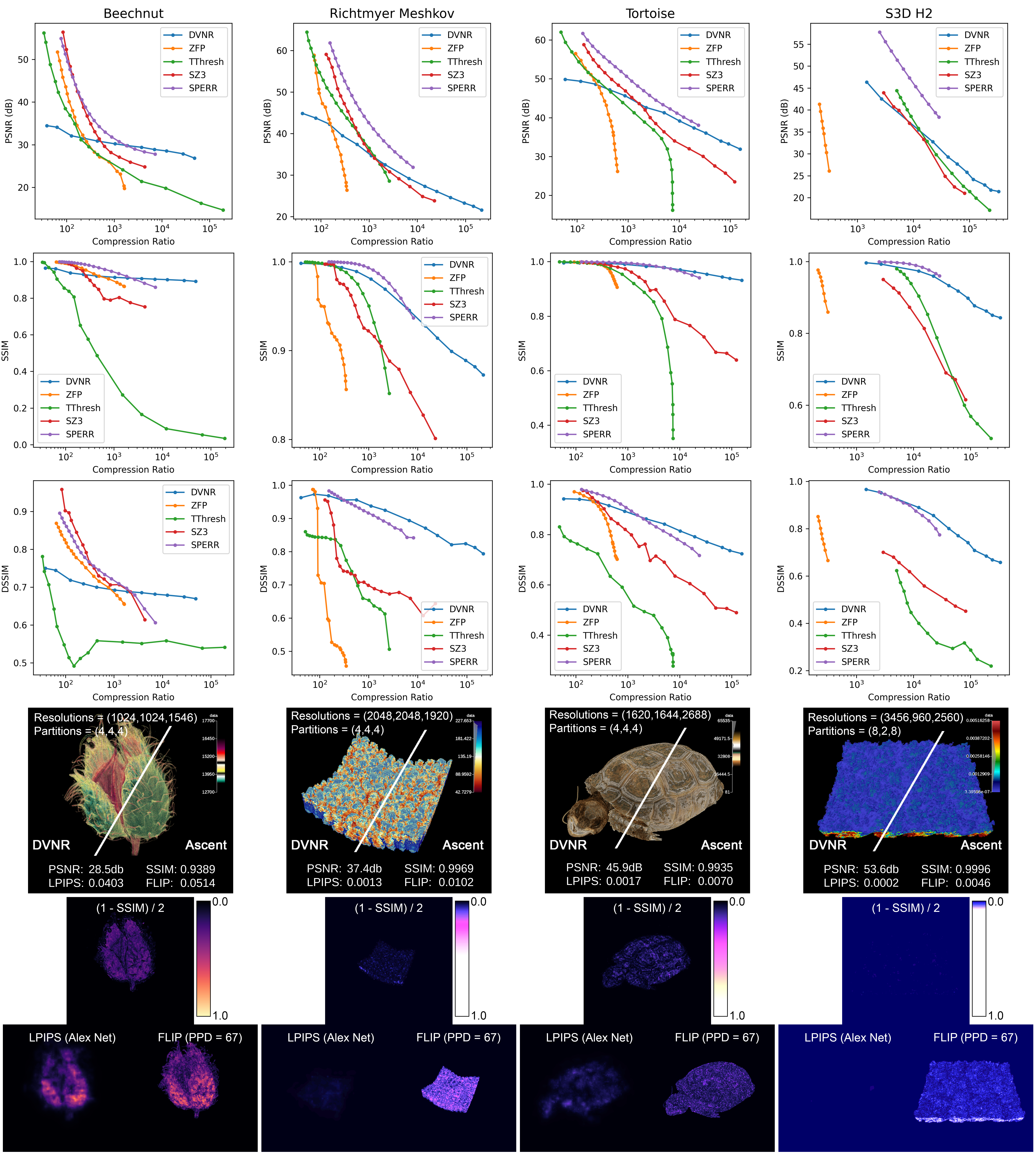}
  \vspace{-2em}
  \caption{Extra results for experiments reported in \Cref{fig:static_volume}.}
  \label{fig:static_volume_more1}
\end{figure}

\onecolumn
\begin{figure}[h]
  \includegraphics[width=\linewidth]{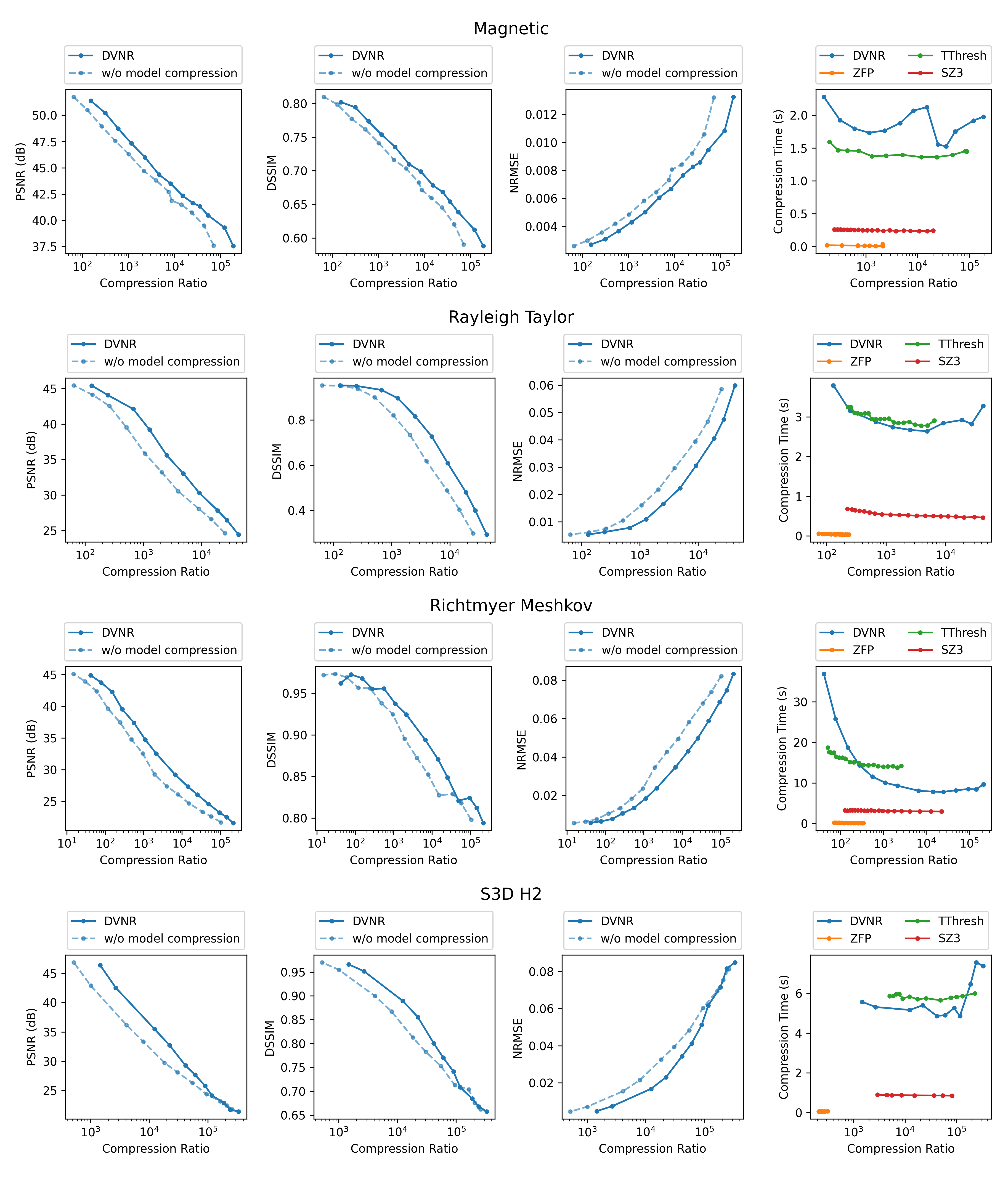}
  \vspace{-2em}
  \caption{Unaveraged results for \Cref{tab:ratios} and \Cref{tab:model_comp_ratios}.}
  \label{fig:unaveraged0}
\end{figure}

\onecolumn
\begin{figure}[h]
  \includegraphics[width=\linewidth]{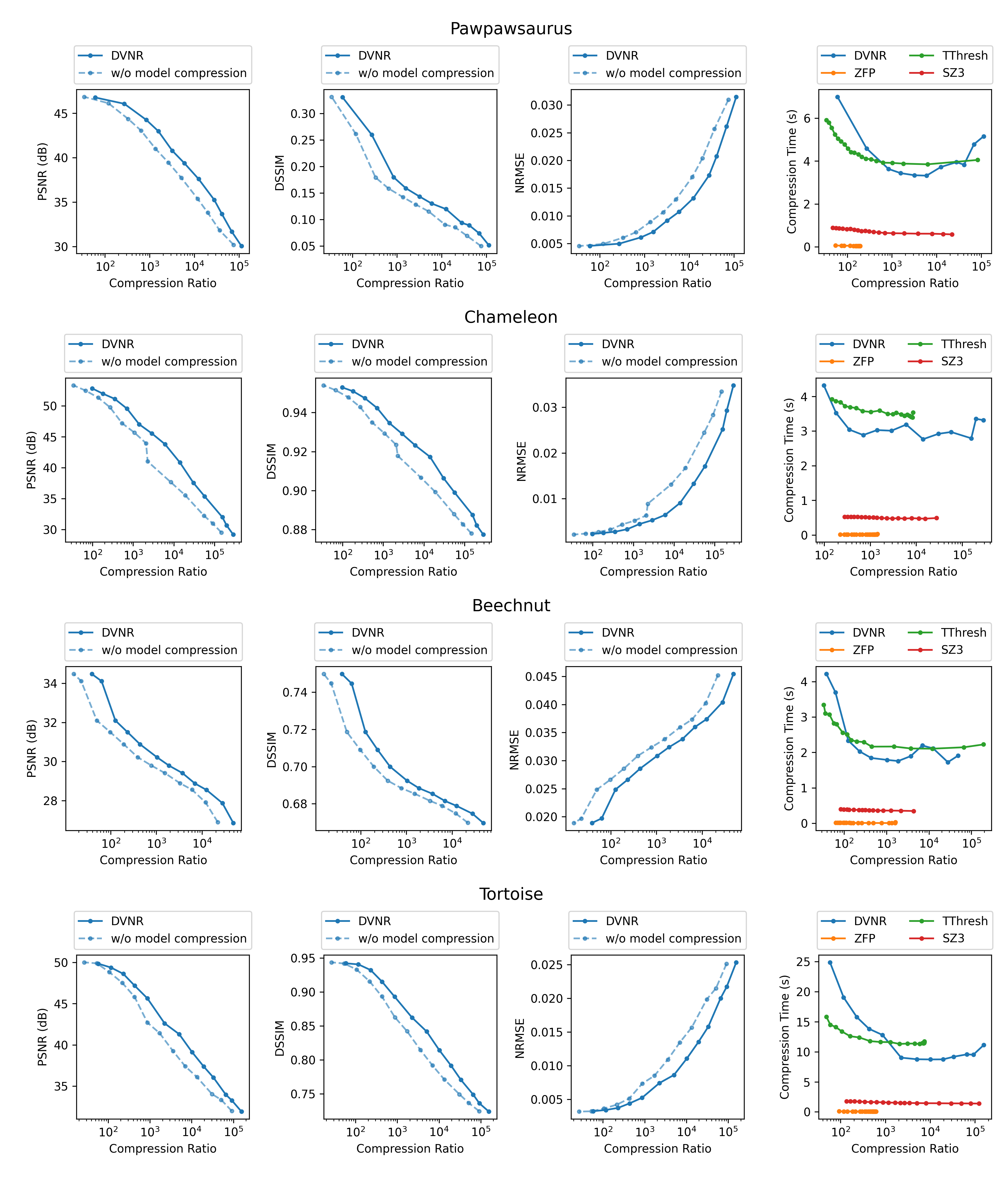}
  \vspace{-2em}
  \caption{Unaveraged results for \Cref{tab:ratios} and \Cref{tab:model_comp_ratios}.}
  \label{fig:unaveraged1}
\end{figure}

\FloatBarrier

\onecolumn
\begin{figure}[h]
  \centering
  \includegraphics[width=\linewidth]{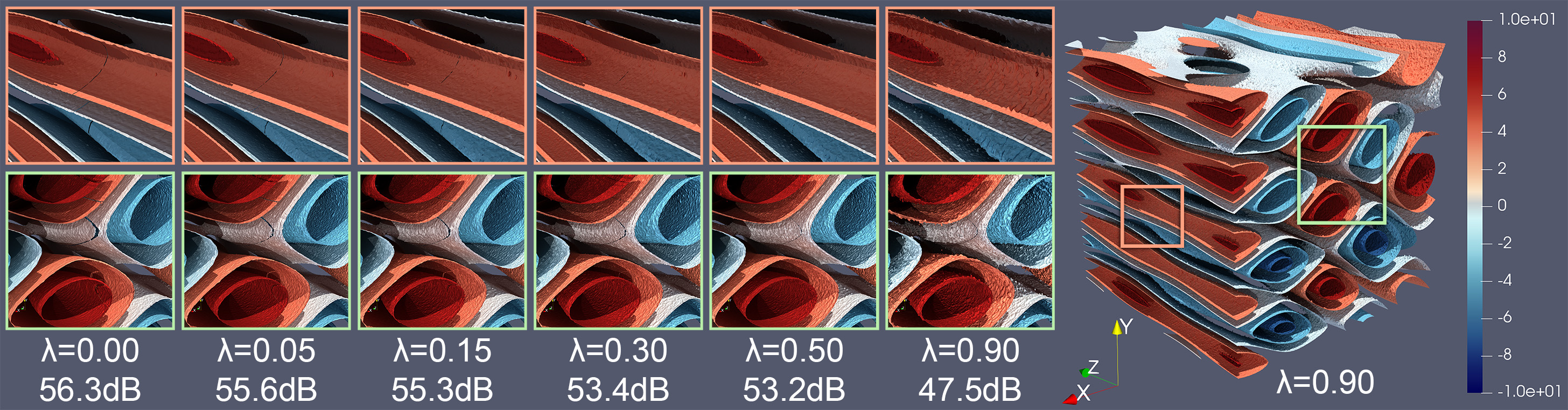}
  \vspace{-2em}
  \caption{\Cref{fig:contour-weights} in high resolution.}
  \vspace{-1em}
\end{figure}

\begin{figure}[h]
  \centering
  \includegraphics[width=\linewidth]{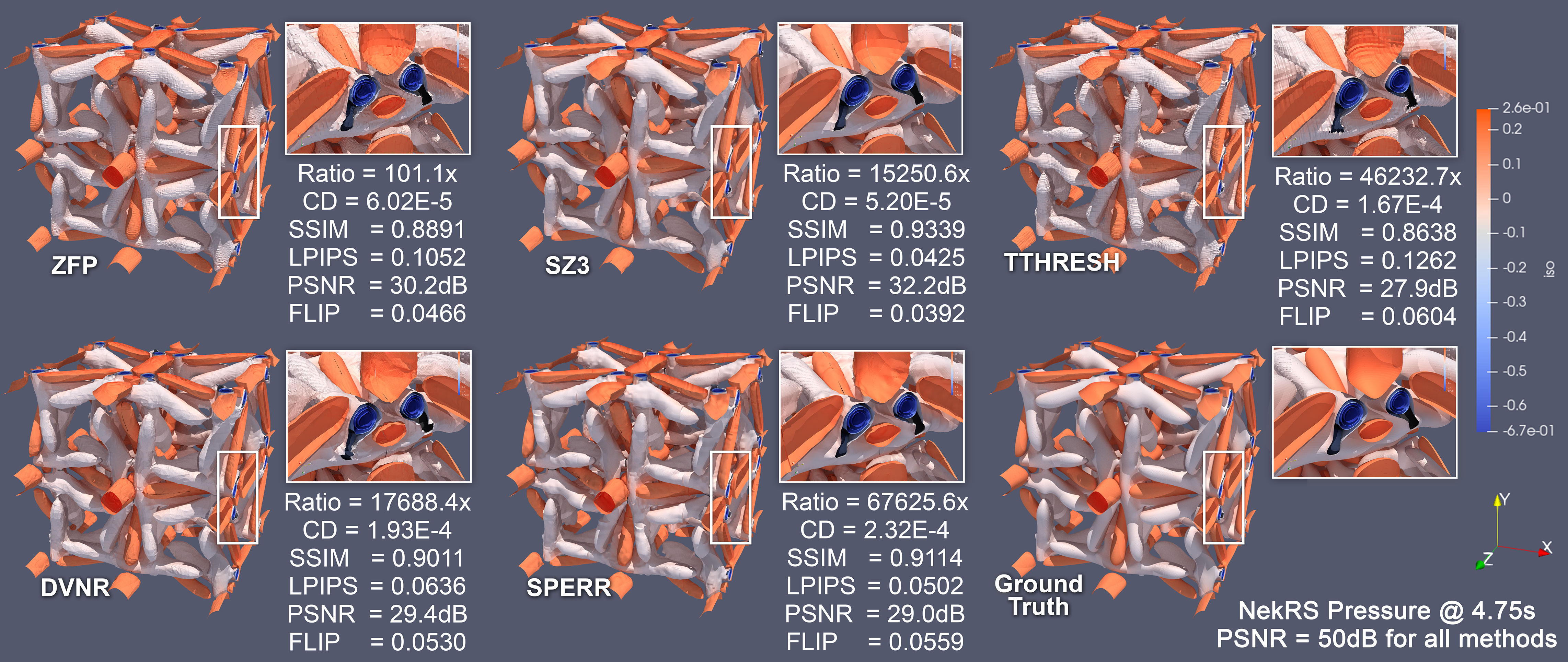}
  \vspace{-2em}
  \caption{\Cref{fig:compression-artifacts} in high resolution.}
  \vspace{-1em}
\end{figure}

\begin{figure}[h]
  \centering
  \includegraphics[width=\linewidth]{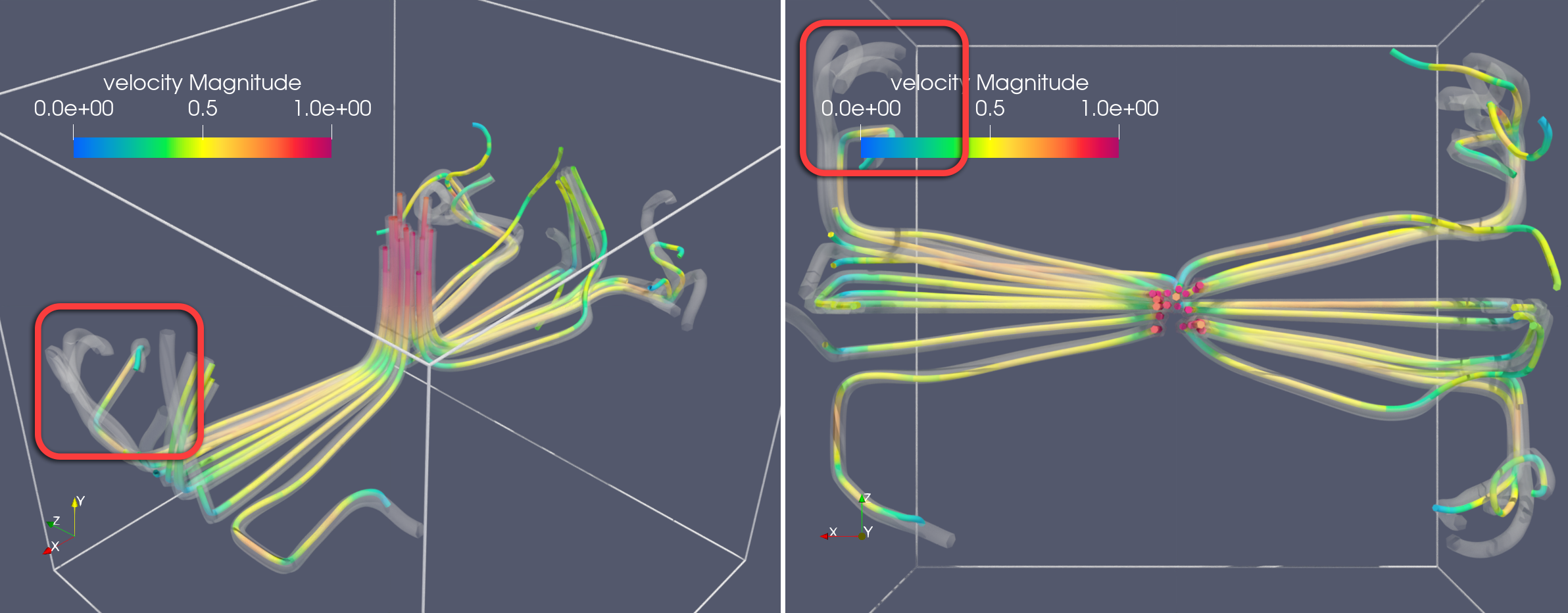}
  \vspace{-2em}
  \caption{\Cref{fig:rendering-pathline} in high resolution.}
  \vspace{-1em}
\end{figure}

\onecolumn
\begin{figure}[h]
  \centering
  \includegraphics[width=\linewidth]{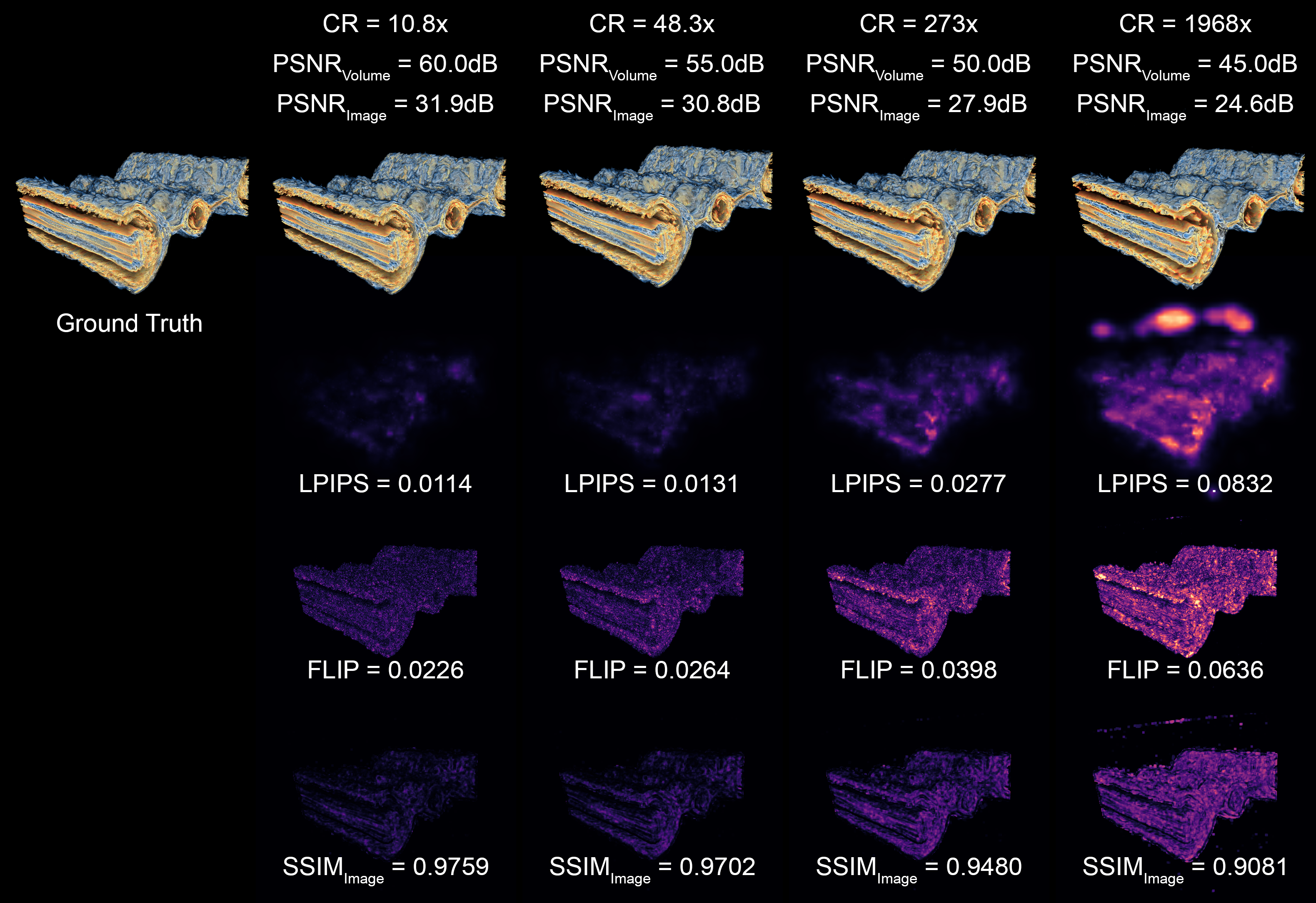}
  \vspace{-2em}
  \caption{Similar to \tvcg{\Cref{fig:rendering_diffpsnr}, with pixel-wise image quality additionally reported.}}
  \label{fig:rendering_diffpsnr_details}
  \vspace{-1em}
\end{figure}

\onecolumn
\begin{figure}[h]
  \centering
  \includegraphics[width=0.75\linewidth]{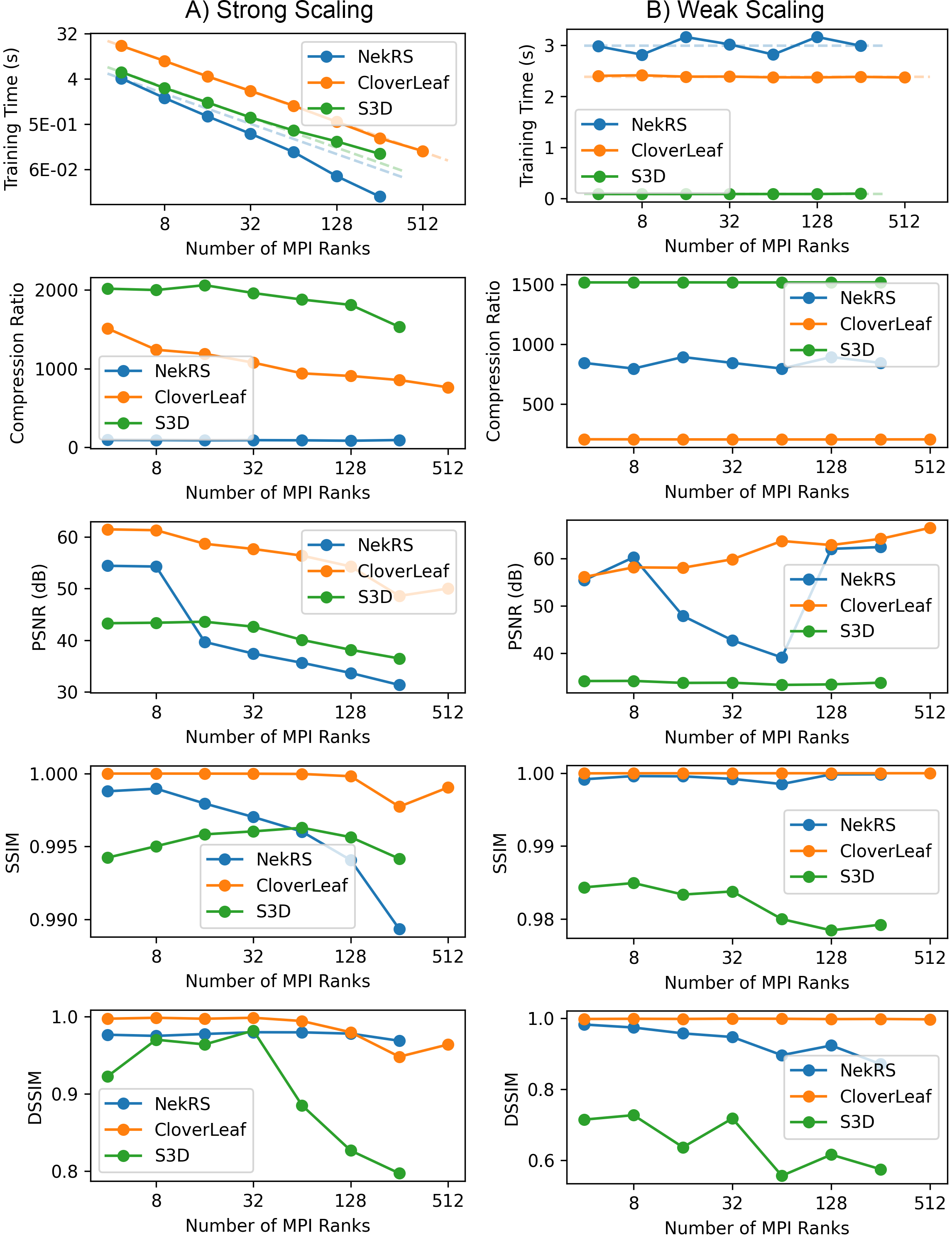}
  \caption{Similar to \Cref{fig:scaling}, with SSIM additionally reported.}
\end{figure}

\onecolumn
\begin{figure}[h]
  \centering
  \includegraphics[width=\textwidth]{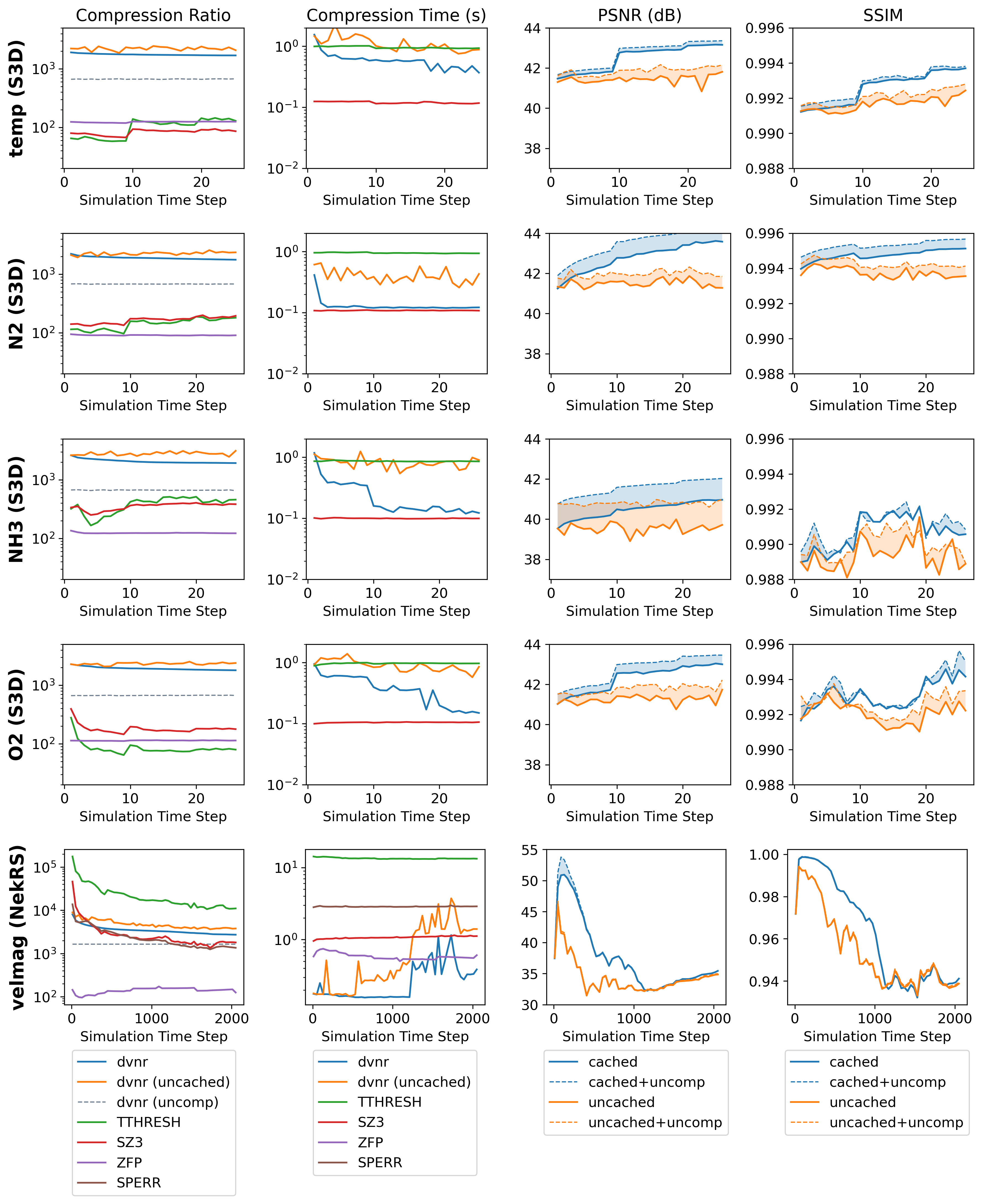}
  \caption{Similar to \Cref{fig:time_evo}, with SSIM additionally reported.}
\end{figure}

\end{document}